%% file: main.tex
\documentclass[a4paper]{jpconf}

\usepackage{preamble}

\begin{document}

\input{macros}

\input{sections/title}

\input{sections/introduction}

\input{sections/section1}

\input{sections/section2}

\input{sections/section3}
\input{sections/section4}

\input{sections/section5}

\input{sections/conclusion}
\input{sections/acknowledgements}

\appendix
\input{sections/appendix1}
\input{sections/appendix2}
\input{sections/appendix3}

\printbibliography

\end{document}

%% file: macros.tex
\definecolor{myblue}{HTML}{20639B}
\definecolor{myblue_darker}{HTML}{173F5F}
\definecolor{mygreen}{HTML}{26A16B}
\definecolor{myyellow}{HTML}{F6D55C}
\definecolor{myred}{HTML}{ED553B}

\newcommand{\Om}[1]{\small $\omega_{#1}$}
\newcommand{\De}[1]{$\Delta_{#1}$}
\newcommand{\Ga}[1]{$\Gamma_{#1}$}

\newcommand{\yb}{\ce{^{171}Yb^+}}
\newcommand{\nbar}{\bar{n}}
\newcommand{\ndot}{\dot{n}}

\newcommand{\vect}[1]{\vv{\bm{#1}}}

%% file: sections/title.tex
\title{Robust entanglement by continuous dynamical decoupling of the J-coupling interaction}

\author{C. H. Valahu$^{1,2}$, A. M. Lawrence$^{1,2}$, S. Weidt$^{1, 3}$, \\ W. K. Hensinger$^{1, 3}$}

\address{$^{1}$Sussex Centre for Quantum Technologies, University of Sussex, Brighton, BN1 9QH, UK}
\address{$^{2}$QOLS, Blackett Laboratory, Imperial College London, London, SW7 2BW, UK}
\address{$^{3}$Universal Quantum Ltd, Brighton, BN1 6SB, UK}

\ead{c.valahu@sussex.ac.uk}

\begin{abstract}
We propose a $\sigma_z\otimes \sigma_z$ laser-free entangling gate which uses the intrinsic J-coupling of ions in a static magnetic gradient. Dephasing of the interaction is suppressed by means of continuous dynamical decoupling using pairs of microwave fields. The gate is virtually insensitive to common amplitude noise of the microwave fields and enables high fidelities despite qubit frequency fluctuations, while the J-coupling interaction's inherent robustness to motional decoherence is retained. Errors far below the fault-tolerant threshold can be achieved at high initial temperatures, negating the requirement of sideband cooling below the Doppler temperature. By adjusting the powers of the continuous microwave fields, the J-coupling interaction can be tuned and can be used to implement parallel entangling gates within an ion chain.
\end{abstract}

\pagebreak

%% file: sections/introduction.tex
\section{Introduction}

Trapped ions have proven to be a promising candidate in the field of Quantum Information Processing (QIP) \cite{bermudez2017}. In the near future Noisy Intermediate-Scale Quantum (NISQ) era, quantum processors will contain dozens of qubits operating in noisy environments \cite{preskill2018}. These qubits must be resilient to noise and robust to imperfect control in order to maintain high fidelities and achieve high circuit depths. Beyond the NISQ-era, Quantum Error Correction (QEC) will lead to error-free QIP, in which many physical qubits encode a single logical qubit. The successful implementation of QEC requires all operation infidelities ($\eta$) to be below the fault-tolerant threshold nearing $\eta < 10^{-2}$ \cite{raussendorf2007}. The number of physical qubits required scales with the infidelity and we therefore define more practical thresholds at $10^{-3}$ and $10^{-4}$. In recent works, laser driven gates have achieved impressive fidelities \cite{ballance2016, gaebler2016, schafer2018, clark2021}, and various dynamical decoupling schemes have been employed to increase robustness \cite{milne2020, shapira2018, biercuk2009, bermudez2012}. Laser-free gates which use microwave and RF radiation are attractive due to their scalability and ease of use, and have shown similar results \cite{mintert2001, ospelkaus2011, harty2016, weidt2016, zarantonello2019, srinivas2021}. Achieving scalable and robust fault-tolerance remains however an important challenge.

\indent An all microwave $\sigma_z\otimes\sigma_z$ type gate has been demonstrated using the intrinsic J-coupling which arises from a string of ions in a static magnetic gradient \cite{khromova2012, piltz2016}. This scheme is attractive because of its robustness to motional decoherence and its minimal overhead due to the passive nature of the interaction. However, since the J-coupling interaction requires magnetic field sensitive states, the gate suffers from spin dephasing and high fidelities are difficult to achieve. Optimized sequences of pulsed dynamical decoupling were shown to extend the coherence time of the spins, although the fidelity remains susceptible to pulse imperfections \cite{piltz2013}.

\indent In this work, we build on Ref. \cite{piltz2016} and propose a $\sigma_z\otimes\sigma_z$ gate using continuous dynamical decoupling which protects the interaction from dephasing. This is similar to Ref. \cite{sutherland2019, srinivas2021} where a laser-free $\sigma_z\otimes\sigma_z$ gate with dynamical decoupling is demonstrated using bichromatic fields. Here, however, we consider only the J-coupling interaction with a static magnetic field gradient. Pairs of microwave fields are applied to each ion which lead to dressed states, a decoherence-free subspace where coherence times similar to clock states have been achieved \cite{timoney2011}. Robustness to motional decoherence is preserved, and the gate is virtually insensitive to amplitude fluctuations of the dressing fields. We show that our dynamical decoupling scheme outperforms existing methods that protect the J-coupling interaction from qubit frequency fluctuations. Furthermore, we show that fidelities beyond the fault-tolerant threshold are possible without ground state cooling and in the presence of motional heating, removing an important bottleneck in QIP. This gate is suitable for fast deployment on NISQ devices due to its simplicity and low requirements. The continuous dressing fields enable a unique controllability of the J-coupling interaction, which leads to useful applications in the field of quantum simulations \cite{cohen2014, cohen2015sim}. Finally, resource-efficient quantum circuits are possible through the parallel execution of entangling gates and the amount of fields only scales with the number of different types of gates in the quantum computer. 

\indent The manuscript is organized in the following manner. The gate Hamiltonian is first derived in section \ref{sec:deriving_hamiltonian} and the gate scheme is explained. Fidelity-damaging terms that were previously neglected are then treated in section \ref{sec:calculating_neglected_terms}. In section \ref{sec:robustness_noise}, the robustness to various noise sources is explored, namely to magnetic field fluctuations, amplitude fluctuations and motional decoherence. In section \ref{sec:fast_gates}, we propose a simple extension to the scheme allowing for faster gates while retaining high fidelities. Finally, in section \ref{sec:applications_and_architectures}, we present our entangling gate in the context of trapped ion QIP and discuss its advantages and applications alongside various existing architectures.

%% file: sections/section1.tex
\section{Deriving the Hamiltonian} \label{sec:deriving_hamiltonian}

In what follows, we consider the trapped ion species \yb. The hyperfine states of the $^2S_{1/2}$ manifold are used as a computational basis and we assign the following labels: $\ket{0} \equiv \ket{F=0, m_f = 0}$ and $\ket{-1,0',+1} \equiv \ket{F = 1, m_f = -1,0,+1}$. Here, $F$ and $m_f$ are the angular momentum and magnetic quantum numbers. The $\ket{\pm 1}$ states are magnetically sensitive while the energy of $\ket{0'}$ is, to first order, magnetic field independent. Due to the long wavelength of microwave and RF radiation, an oscillating or static magnetic field gradient can be introduced to increase spin-motion coupling \cite{mintert2001}. In what follows, we consider only a static gradient. We first introduce the intrinsic J-coupling Hamiltonian and follow the derivation and notation of \cite{wolk2017}. We begin by considering the lab frame Hamiltonian subject to a Polaron transform, $U_p = e^{\frac{1}{2}\epsilon_{j,n}(a - a^\dagger)\sigma_z^{(j)}}$, which decouples the spin and motion (see Refs. \cite{mintert2001, wolk2017, lawrence2019} for more details). After dropping constant terms, the Hamiltonian for a linear string of N ions is

\begin{equation}\label{eq:Hlab}
H = H_{static} + H_J,
\end{equation}
with 
\begin{align}
&	H_{static} = \frac{1}{2} \sum_j \hbar \omega(z_j) \sigma_z^{(j)} + \sum_n \hbar \nu_n \tilde{a}^\dagger_n \tilde{a}_n, \label{eq:H_static} \\
&	H_J = -  \frac{\hbar}{2} \sum_{j<k} J_{j,k}\sigma_z^{(j)}\sigma_z^{(k)}, \label{eq:Hj_lab_frame}
\end{align}

where $\omega(z_j)$ is the frequency of ion $j$ at position $z_j$, $\tilde{a}^\dagger$ ($\tilde{a}$) are the creation (annihilation) phonon operators under the Polaron transform and $\sigma_z$ is the usual Pauli operator defined over the $\{\ket{-1},\ket{+1}\}$ subspace. The magnetic field gradient leads to well defined frequencies $\omega(z_j)$, making individual addressability of ions possible with global MW and RF radiation. The Hamiltonian term $H_J$ describes pairwise couplings of spins in the ion chain. This interaction arises purely from the static magnetic gradient which introduces a position dependent force that allows for conditional logic. While the underlying physical mechanism is different, the spin-spin coupling within an ion chain in a gradient is analogous to the J-coupling present in molecules in the context of NMR \cite{wunderlich2002, wunderlich2003}. The J-coupling constant in equation (\ref{eq:Hj_lab_frame}) is

\begin{equation}
J_{j,k} = \sum_n \epsilon_{j,n}\epsilon_{k,n}  \nu_n,
\end{equation}

which is a summation over all normal modes of vibration $\nu_n$ of the ion string. From here on out, we assume that the magnetic gradient is aligned with the axial $z$ direction of the ion chain and all $\nu_n$ correspond to the axial modes of vibration. Due to the static magnetic gradient, spin and motion are coupled and the effective Lamb-Dicke parameter is \cite{mintert2001}

\begin{equation}
\epsilon_{j,n} = \frac{\mu_B \partial_z B S_{j,n} q_n}{\hbar \nu_n}.
\end{equation}

The magnetic gradient is denoted by $\partial_z B$, $\mu_B$ is the Bohr magneton and $q_n = \sqrt{\hbar/2 m \nu_n}$ is the spatial extent of the ion's wave-function. The elements $S_{j,n}$ indicate how strongly ion $j$ couples to the $n^{th}$ vibrational mode (their values are calculated in Ref. \cite{james1998}). We transform the Hamiltonian $H$ of equation (\ref{eq:Hlab}) into an interaction picture rotating with $H_{static}$, $\tilde{H}_J = e^{\frac{i t}{\hbar}H_{static}}(H - H_{static})e^{\frac{-i t}{\hbar}H_{static}}$, which leads to

\begin{equation}\label{eq:HJint}
\tilde{H}_J = - \frac{\hbar}{2} \sum_{j<k} J_{j,k}\sigma_z^{(j)}\sigma_z^{(k)}.
\end{equation}

The Hamiltonian of equation (\ref{eq:HJint}) forms the basis of the proposal in Refs. \cite{khromova2012, piltz2016}. A $\sigma_z\otimes\sigma_z$ pairwise interaction arises if both qubits are in any combination of the magnetic sensitive states $\ket{\pm 1}$. At durations $\tau = \frac{\pi}{2J}$, a phase of $e^{-i\pi}$ is accumulated and one obtains a controlled phase gate. With additional single qubit rotations, a controlled-NOT gate is implemented. The scheme, however, requires the use of the magnetic sensitive states $\ket{\pm 1}$, which are susceptible to dephasing due to magnetic field fluctuations. The interaction strength can be increased to reduce the gate duration by using a larger magnetic gradient and/or by reducing the secular frequency. Higher gradients, however, are difficult to engineer and may lead to new dominant sources of error. For example, external perturbations which displace the ion chain will lead to larger magnetic field shifts in a bigger gradient. Alternatively, dynamical decoupling pulses are interleaved during the gate to refocus spin states which have undergone decoherence due to low frequency noise. These methods (e.g. CPMG \cite{carr1954, meiboom1958} and Uhrig \cite{uhrig2007}) can increase the coherence time of the sensitive states \cite{piltz2013}, and the best recorded fidelity of such a $\sigma_z\otimes\sigma_z$ gate is 95(3)\% \cite{sriarunothai2018}. The main disadvantage of pulsed dynamical decoupling schemes, however, is that the refocussing pulses themselves are prone to noise and imperfections. The error accumulated by every pulse may impact the fidelity. 

\indent One alternative to pulsed schemes is to continuously drive a transition in order to protect the state from dephasing by opening a new energy gap \cite{timoney2011, webster2013}. Here, we build on Ref. \cite{lawrence2019} and make use of these methods to protect the J-coupling interaction by applying continuous microwave driving fields ($\approx \SI{12.64}{GHz}$) resonant with each of the $\ket{0}\rightarrow\ket{\pm 1}$ transitions. A pair of microwave fields is required per ion since the transition frequencies are well separated in the magnetic field gradient. We consider the Rabi frequencies $\Omega_{dr}$ of all fields to be equal. In the interaction picture of $H_{static}$ and after performing a Rotating Wave Approximation (RWA), the dressing fields are described by \cite{mintert2001}

\begin{equation}\label{eq:Hdr}
\tilde{H}_{dr} = \sum_{j}\frac{\hbar \Omega_{dr}}{2} (\sigma_+^{(j,+)} + \sigma_+^{(j,-)})e^{\sum_n\epsilon_{j,n}(\tilde{a}_n^\dagger - \tilde{a}_n)} + HC,
\end{equation}
 
where $\sigma_+^{(j,\pm)} = \ket{\pm1}^{(j)}\bra{0}^{(j)}$. Hamiltonian (\ref{eq:Hdr}) is expanded in $\epsilon_{j,n}$ up to first order, and higher order terms are neglected in the Lamb-Dicke regime $\epsilon\sqrt{n} \ll 1$ \cite{wineland1998}. The first two terms are

\begin{align}
& \tilde{H}^0_{dr} = \sum_{j}\frac{\hbar\Omega_{dr}}{2} (\sigma_+^{(j,+)} + \sigma_+^{(j,-)}) + HC,\label{eq:Hdr0}\\
& \tilde{H}^1_{dr} =  \sum_{j} \sum_{n}\frac{\hbar \epsilon_{j,n} \Omega_{dr}}{2}(\tilde{a}_n^\dagger e^{ i \nu_n t} - \tilde{a}_n e^{- i \nu_n t})(\sigma_+^{(j,+)} - \sigma_+^{(j,-)}) + HC. \label{eq:Hdr1}
\end{align}

The Hamiltonians (\ref{eq:Hdr0}) and (\ref{eq:Hdr1}) describe carrier and sideband transitions. In the weak driving limit $\Omega_{dr} \ll \nu_n$, $\tilde{H}_{dr}^1$ can be ignored under an RWA. Errors due to this approximation are treated in section \ref{sec:calculating_neglected_terms}. A new eigenbasis arises from the $0^{th}$ order Hamiltonian (\ref{eq:Hdr0}), whose eigenstates and eigenenergies are
\begin{align}
\begin{pmatrix}
\ket{u} \\ \ket{d} \\ \ket{D}
\end{pmatrix} = 
\begin{pmatrix}
\frac{1}{2} & \frac{1}{2} & \frac{1}{\sqrt{2}} \\
\frac{1}{2} & \frac{1}{2} & -\frac{1}{\sqrt{2}} \\
\frac{1}{\sqrt{2}} & - \frac{1}{\sqrt{2}} & 0
\end{pmatrix} 
\begin{pmatrix}
\ket{+1} \\ \ket{-1} \\ \ket{0}
\end{pmatrix} , \ \ \
E =  
\begin{pmatrix}
\Omega_{dr}/\sqrt{2} \\ - \Omega_{dr}/\sqrt{2} \\ 0
\end{pmatrix}.
\end{align}

The eigenbasis forms what is from here on out referred to as the dressed state basis. From the eigenenergies, one can see that the frequency splitting of the dressed states is $\Omega_{dr}/\sqrt{2}$. Therefore, only noise near that frequency will cause population in $\ket{D}$ to leak into the spectator states $\ket{u}$ and $\ket{d}$. The $\ket{D}$ state as well as the spectator states are robust to magnetic field fluctuations and coherence times similar to the clock qubit are achievable \cite{randall2015}. A more rigorous study of the dressed states' robustness is presented in section (\ref{sec:spin_decoherence}). The up and down states $\ket{u}$ and $\ket{d}$ are sensitive to amplitude fluctuations in the microwave drive fields as their energy levels are proportional to $\Omega_{dr}$ and they therefore exhibit worse dephasing times. The dressed states are orthogonal to $\ket{0'}$ and a computational basis can be chosen from $\{\ket{0'}, \ket{D}, \ket{u}, \ket{d}\}$. High fidelity single and two-qubit gates have been demonstrated with the pair $\{\ket{0'}, \ket{D}\}$ \cite{weidt2016}, which are preferred due to their robustness to amplitude fluctuations in the dressing fields.

\indent We now show how to obtain resilient high fidelity gates by combining the spin-spin interaction with the long lived dressed states. The J-coupling Hamiltonian (\ref{eq:HJint}) is transformed to an interaction picture with respect to the dressing fields (\ref{eq:Hdr0}), $e^{i t \tilde{H}^0_{dr}/\hbar} \tilde{H}_J e^{- i t \tilde{H}^0_{dr}/\hbar}$, and in the dressed state basis

\begin{equation}\label{eq:Hj}
\tilde{H}_J = -\hbar J_0(S_+^{(1)}S_+^{(2)} e^{i \sqrt{2}\Omega_{dr} t} + S_-^{(1)}S_-^{(2)} e^{-i \sqrt{2}\Omega_{dr} t} +  S_+^{(1)}S_-^{(2)} +  S_-^{(1)}S_+^{(2)}),
\end{equation}

where we've defined the dressed spin ladder operators $S_+ = \ket{u}\bra{D} + \ket{D}\bra{d}$ and $S_- =\ket{D}\bra{u} + \ket{d}\bra{D}$. We've restricted ourselves to $N=2$ ions for simplicity, set $j=1$, $k=2$ and $J_0 = J_{j=1,k=2}$. In the limit $\Omega_{dr} \gg J_0$, the rotating terms of equation (\ref{eq:Hj}) are neglected under an RWA, leading to

\begin{equation} \label{eq:Hg}
\tilde{H}_G = -\hbar J_0(S_+^{(1)}S_-^{(2)} +  S_-^{(1)}S_+^{(2)}).
\end{equation}

The final expression $\tilde{H}_G$ is the gate Hamiltonian in which both spin operators represent pairwise spin coupling, and qubits are encoded in the subspace $\{\ket{0'}, \ket{D}\}$. The states $\{\ket{0'0'}, \ket{0'D}, \ket{D0'}\}$ are unaffected by the gate Hamiltonian (\ref{eq:Hg}). However, qubit population in the state $\ket{DD}$ is driven out of the computational subspace and transferred to the states $\ket{ud}$ and $\ket{du}$. This is made apparent by solving the time evolution operator $U(t) = e^{-\frac{it}{\hbar}\tilde{H}_G}$ and for the initial state $\ket{\psi_0} = \ket{DD}$,

\begin{equation} \label{eq:ideal_time_evo}
\ket{\psi(t)} = U(t) \ket{\psi(0)} = \cos{(\sqrt{2}J_0t)}\ket{DD} + \frac{i}{\sqrt{2}}\sin{(\sqrt{2}J_0t)} (\ket{ud} + \ket{du}) .
\end{equation}

At durations $\tau = \frac{\pi}{\sqrt{2}J_0}$, the population is transferred back to $\ket{DD}$ and the state has gained a phase of $e^{- i \pi}$. For a two qubit system in the computational subspace $\{\ket{0'0'}, \ket{0'D}, \ket{D0'}, \ket{DD}\}$, this has the overall effect of a controlled-Z gate. Note that the gate duration $\tau$ is $\sqrt{2}$ times slower than the original J-coupling interaction. This gate scheme is unusual in that during the evolution, population leaves the spin's computational subspace $\{\ket{0'}, \ket{D}\}$ but fully returns at the completion of the gate. In section (\ref{sec:robustness_noise}), we show that the two-qubit states $\ket{ud}$ and $\ket{du}$ populated during the evolution remain protected against dephasing and amplitude fluctuations, allowing for high fidelity entanglement. The gate duration remains independent of the dressing power $\Omega_{dr}$ and can be changed by tuning the secular frequency and the magnetic gradient. 

\indent In order to implement two-qubit gates in a controlled way, it is important to be able to disable the J-coupling interaction. This is possible by setting different microwave drive powers for the dressing fields on each ion, i.e. $\Omega^1_{dr} \neq \Omega^2_{dr}$. With this replacement, and after an RWA in the limit $\Omega^1_{dr}, \Omega^2_{dr} \gg J_0$, the gate Hamiltonian becomes

\begin{equation}\label{eq:Hg_rot}
\tilde{H}_G = -\hbar J_0(S_+^{(1)}S_-^{(2)}e^{i t \Delta\Omega^{1,2}/\sqrt{2}} +  S_-^{(1)}S_+^{(2)}e^{-i t \Delta\Omega^{1,2}/\sqrt{2}}),
\end{equation}

where $\Delta\Omega^{1,2} = \Omega^1_{dr} - \Omega^2_{dr}$. By choosing a sufficiently large difference in Rabi frequencies, the terms in equation (\ref{eq:Hg_rot}) can be neglected in an RWA in the limit that $\Delta\Omega^{1,2} \gg J_0$. Thus, pairwise interaction is disabled while preserving the coherence of the dressed states.

%% file: sections/section2.tex
\section{Calculating neglected terms} \label{sec:calculating_neglected_terms}

\indent In the preceding analysis, the gate Hamiltonian was derived and the scheme was explained. However, higher order terms were neglected, namely the first order sideband transition in equation (\ref{eq:Hdr1}) under the assumption that $\Omega_{dr} \ll \nu_n$ and rotating terms in (\ref{eq:Hj}) under an RWA in the limit $\Omega_{dr} \gg J_0$. In this section, we justify these approximations, derive the respective error contributions and show that high fidelities are achievable. 

\indent Infidelities associated with the first order sideband are first analysed. $\tilde{H}_{dr}^1$ (\ref{eq:Hdr1}) is transformed into the interaction picture of $\tilde{H}_{dr}^0$ (\ref{eq:Hdr0}), and expressed in the dressed basis becomes

\begin{equation}\label{eq:Hdr1_I}
\tilde{H}_{dr}^1 = - \sum_{j}\sum_n \frac{\hbar\epsilon_{j, n} \Omega_{dr}}{2} (\tilde{a}_n^\dagger e^{ i \nu_n t} - \tilde{a}_n e^{- i \nu_n t})(S_+^{(j)} e^{i \frac{\Omega_{dr}}{\sqrt{2}}t} - S_-^{(j)} e^{- i \frac{\Omega_{dr}}{\sqrt{2}}t}).
\end{equation}

The Hamiltonian of equation (\ref{eq:Hdr1_I}) is made time independent by using the Magnus Expansion of which a detailed derivation is found in \ref{app:Magnus_expansion}. The leading terms are

\begin{align} \label{eq:Hdr1_ME}
& \tilde{H}_{dr}^1 = \  - \hbar \sum_n \Big[  2\tilde{\mathcal{G}}^1_{1,2, n} \left( S_+^{(1)}S_-^{(2)}  + S_-^{(1)}S_+^{(2)}\right)  \nonumber \\
&\ \ \ \ \ \ \ \ \ \ \ \ \ \ \ \ \ \ \ \ + \sum_j \tilde{\mathcal{G}}_{j,j, n}^1 \left( S_+^{(j)}S_-^{(j)} + S_-^{(j)}S_+^{(j)} \right)  + \tilde{\mathcal{G}}^2_{j, n}  \tilde{a}_n^\dagger\tilde{a}_nS_z^{(j)} \Big] ,  \nonumber \\
 & \tilde{\mathcal{G}}^1_{j,k,n} = \frac{\epsilon_{j,n}\epsilon_{k,n}\Omega_{dr}^2 \nu_n}{2(2\nu_n^2 - \Omega_{dr}^2)},  \ \ \
\tilde{\mathcal{G}}^2_{j,n} = \sum_n\frac{\epsilon_{j,n}^2\Omega_{dr}^3 }{\sqrt{2}(2\nu_n^2 - \Omega_{dr}^2)} . 
\end{align}

Here we've introduced the dressed state operator $S_z = \ket{u}\bra{u} - \ket{d}\bra{d}$. The first terms of (\ref{eq:Hdr1_ME}) are pairwise spin operators identical to the gate Hamiltonian (\ref{eq:Hg}) and contribute to the overall interaction strength. This additional interaction strength does not limit the fidelity as the gate duration can be adjusted accordingly. The second terms are single-ion spin operators which contain unwanted transitions causing the qubits to remain entangled with spectator dressed states at the completion of the gate. The final term leads to static shifts of the dressed state by an energy proportional to the thermal state. The $S_z$ operator commutes, however, with all other dressed state operators and can therefore be omitted as it has no influence on the gate. 

 An approximate time-independent Hamiltonian is defined from equations (\ref{eq:Hj}) and (\ref{eq:Hdr1_ME}),

\begin{equation}\label{eq:Hg_and_Hsingle}
\tilde{H} = \tilde{H}_G + \tilde{H}_{single} = - \hbar J_{tot} \big( S_+^{(1)}S_-^{(2)}  + S_-^{(1)}S_+^{(2)} \big) - \hbar \sum_n\sum_j \tilde{\mathcal{G}}^1_{j,j,n}( S_+^{(j)}S_-^{(j)} + S_-^{(j)}S_+^{(j)} ).
\end{equation}

The modified J-coupling interaction strength is $J_{tot} = J_0 + J_{eff}$, where $J_{eff} = \sum_n 2\tilde{\mathcal{G}}^1_{1,2,n}$. Errors due to the single-ion spin terms of $\tilde{H}_{single}$ are best understood by expanding the dressed spin operators, such that

\begin{equation} \label{eq:Hsingle}
\tilde{H}_{single} = - \hbar \sum_n\sum_j\tilde{\mathcal{G}}^1_{j,j,n}  (2\ket{D}\bra{D}^{(j)} + \ket{u}\bra{u}^{(j)} + \ket{d}\bra{d}^{(j)}).
\end{equation}

\begin{figure}[t]
\centering
\includegraphics[scale=1.1]{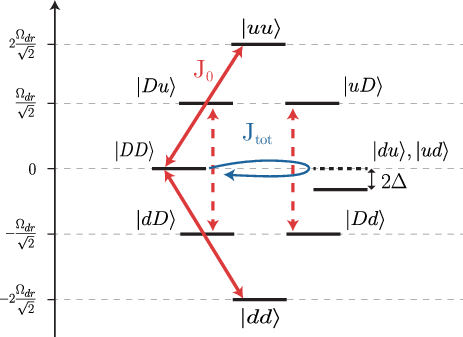} 
\caption{Two qubit energy level diagram, obtained by moving the $0^{th}$ order dressing Hamiltonian (\ref{eq:Hdr0}) into the dressed state basis, i.e. $\tilde{H}^0_{dr} = \frac{\hbar\Omega_{dr}}{\sqrt{2}}\sum_j S_z^{(j)}$. The gate's main transition is represented by the blue arrow with a corresponding interaction strength $J_{tot}$. The single ion terms of $H_{single}$ (equation (\ref{eq:Hsingle})) make the interaction off-resonant by an amount $2\Delta$. The red arrows illustrate the four possible pairwise spin transitions arising from Hamiltonian (\ref{eq:Hrot_dressed}), $H_{rot}$, each with strength $J_0$. Dashed arrows show transitions which do not affect the gate's computational subspace. }
\label{fig:energy_diagram}
\end{figure}

The energy of the $\ket{D}^{(j)}$ state is shifted  by $\hbar\Delta_j = \hbar\sum_n\tilde{\mathcal{G}}^1_{j,j,n}$ relative to the $\ket{u}^{(j)}$ and $\ket{d}^{(j)}$ states. This causes the J-coupling interaction to be off-resonant and, for $N=2$ ions where $\Delta_1 = \Delta_2 \equiv \Delta$, introduces a detuning of $2\Delta$. The J-coupling interaction can therefore be thought of as a virtual field interacting with the $\ket{DD} \rightarrow \frac{1}{\sqrt{2}}(\ket{ud} + \ket{du})$ transition and is made off-resonant due to terms in $\tilde{H}_{single}$ (see figure (\ref{fig:energy_diagram})).

The modified time evolution $\ket{\psi(t)} = e^{- i t \tilde{H}/\hbar}\ket{DD}$ obtained from (\ref{eq:Hg_and_Hsingle}) becomes

\begin{equation} \label{eq:evo_psi_jdelta}
\ket{\psi(t)} =  e^{3 i \Delta  t}\left[\Big(\cos{(J_\delta t)} + \frac{i\Delta}{J_\delta}\sin{(J_\delta t)}\Big)\ket{DD} + \frac{i J_{tot}}{J_\delta}\sin{(J_\delta t)} \left(\ket{ud} + \ket{du}\right) \right],
\end{equation}

where we've defined $J_\delta = \sqrt{\Delta^2 +2  J_{tot}^2}$. We again draw a parallel between the J-coupling interaction and a virtual field as the expression for $J_\delta$ is identical to the Rabi frequency of a detuned field. A new gate duration $\tau =  \pi/J_\delta$ is defined from (\ref{eq:evo_psi_jdelta}) after which all population is returned to the initial state $\ket{DD}$. An additional phase $e^{3i\phi}$ is acquired, where $\phi = \pi \Delta /J_\delta$. Although small, this parasitic phase may spoil the performance of the gate. We calculate the fidelity of obtaining the Bell state $\ket{\Psi^-}$, which is identical to calculating the overlap between the ideal entangled target state $\ket{\Psi}_T = \frac{1}{2}\left(\ket{0'0'} + \ket{0'D} + \ket{D0'} - \ket{DD}\right)$ with the resulting state $ U \ket{\Psi}_0 = e^{-\frac{i \tau}{\hbar} \tilde{H}}\frac{1}{2}( \ket{0'0'} + \ket{0'D} + \ket{D0'} + \ket{DD})$, such that 

\begin{equation} \label{eq:fidelity}
\mathcal{F} = |\bra{\Psi}_T U \ket{\Psi}_0 |^2.
\end{equation}

This expression is used to compare the fidelity of the entangling gate with respect to the fault-tolerant threshold throughout the remainder of the text. This comparison may appear misleading when considering the error processes that contribute to the infidelity. In the following derivations, it is seen that an intrinsic error mechanism leaks population out of the computational subspace. Furthermore, incoherent leakage may also be caused by dephasing noise. Leakage errors can not traditionally be corrected for by error-correction protocols. However, several works have mapped leakage into error channels that are compatible with error-correction codes \cite{aliferis2007, brown2019, stricker2020}. Furthermore, leakage within the hyperfine ground state of \yb can be transformed into error channels compatible with surface codes with no additional circuit complexity \cite{hayes2020} . In this way, the aforementioned fault-tolerant threshold remains a useful metric of comparison.

Under the time evolution operator of Hamiltonian (\ref{eq:Hg_and_Hsingle}), the initial state $\ket{\Psi}_0$ becomes

\begin{equation}
U\ket{\Psi}_0 = \frac{1}{2} \left[\ket{0'0'} + e^{2i\phi}\left(\ket{0'D} + \ket{D0'}\right) - e^{3i\phi}\ket{DD}\right],
\end{equation}

and the fidelity is

\begin{equation} \label{eq:fidelity_1}
\mathcal{F} = \frac{1}{8}\left(3 + 2\cos{(\phi)} + 2\cos{(2\phi)} + \cos{(3\phi)}\right).
\end{equation}

An approximate infidelity $\eta_1 = 1-\mathcal{F}$ is found after identifying the leading terms from the Taylor expansion of (\ref{eq:fidelity_1}),

\begin{equation} \label{eq:eta_1}
\eta_1 = \frac{475 \pi^2 \Omega_{dr}^4}{4608\nu_1^4}.
\end{equation}

From equation (\ref{eq:eta_1}), we confirm that infidelities arising from off-resonant coupling of the carrier to the motional sidebands can be suppressed by decreasing the dressing fields' Rabi frequency $\Omega_{dr}$ or increasing the vibrational frequency $\nu_1$.

We now move to terms rotating in the J-coupling Hamiltonian of equation (\ref{eq:Hj}) which were neglected under an RWA in the limit $\Omega_{dr} \gg J_0$. This approximation does not always hold for realistic parameters and the RWA may break down. After expanding the dressed state spin operators, the rotating terms of (\ref{eq:Hj}) become

\begin{equation}\label{eq:Hrot_dressed}
H_{rot} = -\hbar J_0(\ket{DD}\bra{dd} + \ket{uu}\bra{DD} + \ket{Du}\bra{dD} + \ket{uD}\bra{Dd})e^{i\sqrt{2}\Omega_{dr} t} + \textrm{HC}.
\end{equation}

The Hamiltonian $H_{rot}$ describes transitions induced by off-resonant coupling of the virtual J-coupling interaction field to spectator dressed states separated by $\sqrt{2}\Omega_{dr}$. The $\ket{Du}\leftrightarrow\ket{dD}$ and $\ket{uD}\leftrightarrow\ket{Dd}$ transitions have no effect on the gate as these states are never populated. Population may however leak out of the gate's subspace through the $\ket{DD}\leftrightarrow\ket{uu}$ and $\ket{DD}\leftrightarrow\ket{dd}$ transitions (see figure (\ref{fig:energy_diagram})). In order to derive an expression for the infidelity, $H_{rot}$ is made time-independent by moving into an interaction picture with respect to $H_{shift}$, such that $\tilde{H}_{rot} = e^{i t H_{shift}}(H_{rot} - H_{shift})e^{-itH_{shift}}$, where

\begin{equation} \label{eq:Hshift}
H_{shift} = - \frac{\hbar \Omega_{dr}}{\sqrt{2}}\sum_j S_z^{(j)},
\end{equation}

resulting in 

\begin{align}\label{eq:Hrot_int}
\tilde{H}_{rot} = & -\hbar J_0(S_+^{(1)}S_+^{(2)} + S_-^{(1)}S_-^{(2)}) - H_{shift}.
\end{align}

\begin{figure}[t]
\centering
\includegraphics[scale=1]{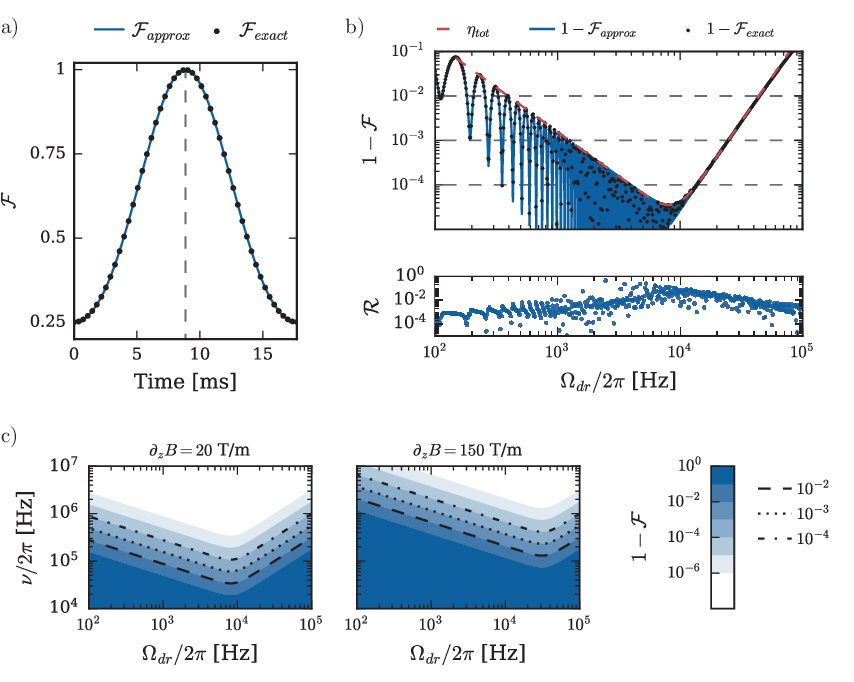} 
\caption{Simulated and predicted infidelities of the J-coupling interaction. (a) Time evolution of the fidelity with $N=2$ ions, $\Omega_{dr}/2\pi = \SI{20}{kHz}$, $\nu/2\pi= \SI{140}{kHz}$ and $\partial_zB=\SI{20}{T/m}$, resulting in $J_0/2\pi = \SI{39.4}{Hz}$. The maximum is reached at $\tau = \pi/\sqrt{2}J_{tot} = \SI{8.86}{ms} $. The approximate infidelities $\mathcal{F}_{approx}$ are obtained by numerically integrating $H_{approx}$ (\ref{eq:Happrox}), while $\mathcal{F}_{exact}$ originates from $H_{exact}$ (\ref{eq:Hexact}). The Hilbert space of both vibrational modes is truncated at $N_{fock} = 3$ and the initial temperature is $n_{th} = 0$. (b) The fidelity at the gate duration is plotted for a range of Rabi frequencies (top). The same parameters as in (a) are used. The total analytical infidelity $\eta_{tot}$ from equation (\ref{eq:eta_tot}) is plotted in red. (bottom) Fractional residuals of exact and approximate fidelities calculated as $\mathcal{R} = |\mathcal{F}_{approx} - \mathcal{F}_{exact}|/(1 - \mathcal{F}_{exact})$. (c) Expected infidelities from $\eta_{tot}$ (\ref{eq:eta_tot}) for varying dressing powers $\Omega_{dr}$ and secular frequencies $\nu$. Here the label represents the first mode $\nu_1$, and the second mode's frequency can be inferred from $\nu_2 = \sqrt{3}\nu_1$. A gradient of $\SI{20}{T/m}$ ($\SI{150}{T/m}$) is considered in the left (right) plot.}
\label{fig:error_analysis}
\end{figure}

Here we've omitted the $\ket{Du}\leftrightarrow\ket{dD}$ and $\ket{uD}\leftrightarrow\ket{Dd}$ terms as they have no effect on the gate. The total Hamiltonian becomes

\begin{equation}\label{eq:Hj_time_ind}
\tilde{H} = \tilde{H}_G + \tilde{H}_{rot},
\end{equation}

where $\tilde{H}_G$ refers to the ideal gate Hamiltonian of (\ref{eq:Hg}). $\tilde{H}_G$ and $H_{shift}$ commute with one another, therefore the gate Hamiltonian remains unaffected in the interaction picture with respect to $H_{shift}$. Since Hamiltonian (\ref{eq:Hj_time_ind}) does not affect the states $\ket{0'D}$, $\ket{D0'}$ and $\ket{0'0'}$, it is sufficient to only derive the time evolution of $\ket{\psi}_0 = \ket{DD}$ to evaluate the fidelity. At the gate duration $\tau = \frac{\pi}{\sqrt{2}J_0}$, the amplitude $a$ of the state $\ket{DD}$  is

\begin{align} \label{eq:amp_dd_state}
& a_{DD} =  \frac{1}{2} ( \beta_1 + \beta_2 ) + \frac{2J_0^2 - \Omega_{dr}^2}{2\sqrt{4J_0^4 + \Omega_{dr}^4}} ( \beta_1 - \beta_2 ),\\
& \beta_1 = \cos{(\frac{\pi\sqrt{2J_0^2 + \Omega_{dr}^2 + \sqrt{4J_0^4 + \Omega_{dr}^4}}}{\sqrt{2}J_0})},\\
& \beta_2 = \cos{(\frac{\pi\sqrt{2J_0^2 + \Omega_{dr}^2 - \sqrt{4J_0^4 + \Omega_{dr}^4}}}{\sqrt{2}J_0})}.
\end{align}

In the limit $\Omega_{dr} \gg J_0$, the expected amplitude $a_{DD} = -1$ is retrieved. For higher J-coupling strengths, the state may not fully end up in $\ket{DD}$ at the completion of the gate, as population remains in $\ket{uu}$ and $\ket{dd}$. The $\beta_1$ term leads to oscillations of the population while $\beta_2\rightarrow -1$ for large $\Omega_{dr}$. An upper bound on the infidelity is therefore estimated by setting $\beta_1\rightarrow 1$ and with the fidelity equation defined in (\ref{eq:fidelity}), the error becomes

\begin{equation} \label{eq:eta_2}
\eta_2 = \frac{J_0^2}{\Omega_{dr}^2}.
\end{equation}

By decreasing $J_0$, the power of the virtual field is lessened and population transfer to spectator states due to off-resonant coupling is mitigated. Alternatively, one can minimize the infidelity of equation (\ref{eq:eta_2}) with a greater Rabi frequency $\Omega_{dr}$ which increases the energy gap between $\ket{DD}$ and $\ket{uu}$, $\ket{dd}$.

We now summarize the aforementioned error sources. We first consider the J-coupling interaction as a virtual field of strength $J_{tot}$ which couples the main gate transition $\ket{DD} \rightarrow (\ket{ud} + \ket{du})/\sqrt{2}$. Large Rabi frequencies of the dressing field $\Omega_{dr}$ may off-resonantly couple to the motional sidebands. In the dressed interaction picture, this is analogous to introducing an energy shift of the $\ket{ud}$ and $\ket{du}$ states, causing the virtual gate field to be off-resonant. An approximate error term is given in (\ref{eq:eta_1}) and can be minimized by ensuring that $\Omega_{dr} \ll \nu_1$. Conversely, smaller Rabi frequencies may cause the virtual gate field to off-resonantly couple to spectator dressed states separated in frequency by $\sqrt{2}\Omega_{dr}$. This causes population to leave the computational subspace and the approximate error in equation (\ref{eq:eta_2}) is negligible in the limit $\Omega_{dr} \gg J_0$. The gate transition and intrinsic error mechanisms in the dressed interaction picture are illustrated in figure (\ref{fig:energy_diagram}). Both error terms are combined and represent the total intrinsic infidelity of our gate scheme,

\begin{equation} \label{eq:eta_tot}
\eta_{tot} = \eta_1 + \eta_2.
\end{equation}

The veracity of the Magnus Expansion and the error analysis leading up to equation (\ref{eq:eta_tot}) is verified by numerically integrating the exact Hamiltonian made up of (\ref{eq:HJint}), (\ref{eq:Hdr0}) and (\ref{eq:Hdr1}),

\begin{equation} \label{eq:Hexact}
H_{exact} = \tilde{H}_J + \tilde{H}^0_{dr} + \tilde{H}^1_{dr},
\end{equation}

and comparing it to the approximate time-independent Hamiltonian\footnote{
The final Hamiltonian $H_{approx}$ is the result of three successive interaction pictures that offer insights into the various infidelity mechanisms. We first move into the ion frame rotating with $H_{static}$ of equation (\ref{eq:H_static}). The J-coupling and first order dressing Hamiltonian (equations (\ref{eq:HJint}) and (\ref{eq:Hdr1})) are then transformed to an interaction picture with respect to the zeroth order dressing Hamiltonian (\ref{eq:Hdr0}). The final terms of $H_{approx}$ lie in an interaction picture rotating with $H_{shift}$ (\ref{eq:Hshift}). Note that both $\tilde{H}_{G}$ and $\tilde{H}_{rot}$ commute with $H_{shift}$ and therefore remain unaffected.} obtained from (\ref{eq:Hg_and_Hsingle}), (\ref{eq:Hshift}) and (\ref{eq:Hrot_int}),

\begin{equation} \label{eq:Happrox}
H_{approx} = \tilde{H}_G + \tilde{H}_{single} + \tilde{H}_{rot},
\end{equation}

as well as the total intrinsic infidelity $\eta_{tot}$. The time evolution of the fidelity is shown in figure (\ref{fig:error_analysis}a). The dependence of the fidelity on the Rabi frequency $\Omega_{dr}$ is plotted in figure (\ref{fig:error_analysis}b) and one can see that the approximate and exact Hamiltonians are in good agreement. Furthermore, the analytical error model $\eta_{tot}$ is a good fit to the exact infidelities. For smaller $\Omega_{dr}$, $\eta_{tot}$ is an envelope to the highly oscillatory behaviour of the exact errors. In this regime, errors are dominated by off-resonant coupling of the virtual J-coupling interaction field to spectator dressed states, and we verify the quadratic scaling $1/\Omega^{2}_{dr}$ as indicated by equation (\ref{eq:eta_2}). For higher powers, we observe a $\Omega_{dr}^4$ scaling from equation (\ref{eq:eta_1}) which arises from off-resonant coupling of the physical dressing fields to the motional sidebands. Despite the intrinsic error mechanisms, fidelities well below the fault tolerant threshold $<10^{-2}$ are possible for a wide range of dressing powers. The overall infidelity can be minimized by appropriately choosing $\Omega_{dr}$. The optimal dressing Rabi frequency is found from (\ref{eq:eta_tot}),

\begin{equation} \label{eq:Omega_dr_opt}
\Omega_{dr}^{opt} = 2\left(\frac{\partial_zB^2 \mu_B^2}{5\sqrt{19}\hbar m \pi}\right)^{1/3}.
\end{equation}

For a gradient of $\SI{20}{T/m}$ one finds $\Omega_{dr}^{opt}/2\pi = \SI{8.1}{kHz}$ which is well within experimental capabilities. Note that $\Omega^{opt}_{dr}$ is independent of the motional frequency. 

\begin{figure}[t]
\centering
\includegraphics[scale=1]{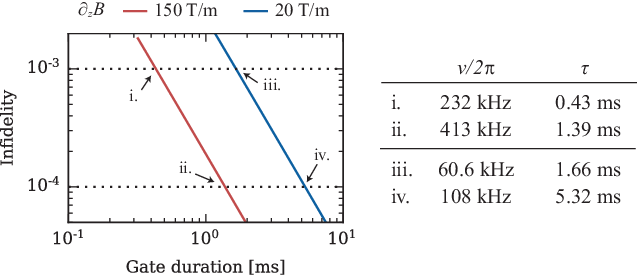} 
\caption{Minimum achievable infidelities for various gate durations. The infidelities are calculated from (\protect\ref{eq:eta_tot}) after setting the dressing Rabi frequency to $\Omega_{dr}^{opt}$ (\ref{eq:Omega_dr_opt}). We consider two magnetic field gradient strengths of $\SI{20}{T/m}$ and $\SI{150}{T/m}$ with respective Rabi frequencies $\SI{8.1}{kHz}$ and $\SI{31.2}{kHz}$. Pairs of secular frequencies and gate durations are provided for points indicated by the annotations i-iv.}
\label{fig:error_vs_gate_duration}
\end{figure}

In figure (\ref{fig:error_analysis}c), the infidelity $\eta_{tot}$ is plotted for a range of secular frequencies and dressing powers, as well as for gradients $\partial_zB = 20$ and $\SI{150}{T/m}$. The inherently slow speed of the J-coupling interaction may be counteracted by lowering the motional frequency. However, from figure (\ref{fig:error_analysis}c), smaller motional frequencies lead to higher infidelities. Indeed, both error terms in $\eta_{tot}$ increase with smaller $\nu_n$: on the one hand, off-resonant coupling to the sidebands increases as the frequency gap set by $\nu_n$ decreases, and on the other hand the J-coupling interaction strength increases resulting in more off-resonant coupling to the spectator dressed states. This ultimately limits the achievable speed of the interaction and introduces a trade-off between the gate duration and the fidelity (see figure (\ref{fig:error_vs_gate_duration})). The motional frequency and dressing powers can be chosen to obtain a gate with errors below a certain threshold, and with these considerations the J-coupling interaction strength scales as $J_0 \propto \partial_zB^{3/2}$. Notice that the interaction strength is now independent of the secular frequency and only the gradient can be changed to decrease the gate duration. As an example, in order to achieve errors $< 10^{-4}$ with a gradient of $\SI{150}{T/m}$, a secular frequency $\nu_1/2\pi = \SI{412.7}{kHz}$ is required resulting in an interaction strength $J_0/2\pi = \SI{254.9}{Hz}$ and a gate duration $\tau = \SI{1.39}{ms}$. Despite a larger gradient, the gate duration is relatively slow. In section 5, we show how to protect the interaction against errors arising from lower motional frequencies which allows the gate speed to increase by almost an order of magnitude while achieving the same fidelity.

We finally consider the off-resonant interactions of the dressing fields to the transitions of another ion. The static magnetic field gradient leads to individual addressability of the ions in frequency space. The $\ket{0}\rightarrow \ket{\pm 1}$ transitions of neighbouring ions are separated by several MHz. Large dressing field Rabi frequencies may therefore off-resonantly drive the transition of another ion, as well as change the energy level through an AC stark shift. At a secular frequency of $\SI{108}{kHz}$ ($\SI{413}{kHz}$) with a gradient of $\SI{20}{T/m}$ ($\SI{150}{T/m}$), the frequency splitting between two ions is $\SI{4.3}{MHz}$ ($\SI{13.1}{MHz}$). The off-resonant excitation probability is estimated to be smaller than $\Omega_{dr}^2/(\Omega_{dr}^2 + \delta^2)$ \cite{khromova2012}, which, at the optimal dressing Rabi frequency, is less than $\num{3.5e-6}$ ($\num{5.6e-6}$). Errors due to off-resonant drivings of another ion are therefore negligible in front of the infidelity terms of equation (\ref{eq:eta_tot}). They remain negligible at higher Rabi frequencies since the excitation probability is proportional to $\Omega_{dr}^2$ whereas infidelities from coupling to the motional sidebands scale with $\Omega_{dr}^4$ (see equation (\ref{eq:eta_1})). The effects of AC stark shifts were verified by numerically simulating the exact Hamiltonian (\ref{eq:Hexact}) along with static energy shifts. At the optimal dressing Rabi frequencies, the added infidelity is negligible ($\approx 10^{-8}$). At larger Rabi frequencies, the errors are again dominated by off-resonant coupling to the motional sidebands.

%% file: sections/section3.tex
\section{Robustness to noise sources} \label{sec:robustness_noise}~\\
Here we consider several noise sources and their contributions to the gate's infidelity. We first study noise which affects the spin's coherence, namely magnetic field fluctuations which cause dephasing. Another contribution is fluctuations in the driving field's Rabi frequency. We then consider motional decoherence and show robustness to thermal noise and motional heating.

\subsection{Spin Decoherence} \label{sec:spin_decoherence}

\subsubsection*{Magnetic field fluctuations}\hspace*{\fill} \\

We model dephasing as fluctuations of the qubit frequency,

\begin{align} \label{eq:Hnoise}
H_{noise} = \sum_j \sum_{m} \frac{ \delta\omega_m(t)}{2}  \sigma_z^{(j,m)},
\end{align}

where $m$ indexes over the $\ket{\pm 1, 0'}$ states. To good approximation for \yb, noise in the qubit frequency is linear with small magnetic field fluctuations, $\delta\omega_m(t) = \delta B(t) \partial_B\omega_m$, where $\partial_B\omega_m$ are sensitivities of the states $\ket{m}$. Since the $\ket{\pm 1}$ states are sensitive to first order to magnetic field noise, $\partial_B\omega_{\pm 1} \gg \partial_B\omega_{0'}$. Equation (\ref{eq:Hnoise}) is transformed into the dressed interaction picture with $H_{dr}^0$ of equation (\ref{eq:Hdr0}), $e^{i t H^0_{dr}/\hbar} H_{noise} e^{- i t H^0_{dr}/\hbar}$, resulting in

\begin{align} \label{eq:Hnoise_int}
& \tilde{H}_{noise} = \tilde{H}_{fluc} + \tilde{H}_{leak}, \\
& \tilde{H}_{fluc}  = \sum_j \frac{\delta B(t)}{8} \partial_B\omega_{0'} \Big( 4 \ket{0'}\bra{0'} + 2 \ket{D}\bra{D} - 3(\ket{d}\bra{d} + \ket{u}\bra{u})\Big), \label{eq:Hfluc}\\
& \tilde{H}_{leak}  = \sum_j \frac{\delta B(t)}{4\sqrt{2}}(\partial_B\omega_{+1} - \partial_B\omega_{-1})(S_+^{(j)} e^{i t \Omega/\sqrt{2}} + S_-^{(j)} e^{-i t \Omega/\sqrt{2}}) \nonumber \\
 & + \frac{5\delta B(t)}{8}\partial_B\omega_{0'}(\ket{u}\bra{d} e^{2i t \Omega/\sqrt{2}} + \ket{d}\bra{u} e^{-2i t \Omega/\sqrt{2}}).  \label{eq:Hleak}
\end{align}

The noise Hamiltonian $\tilde{H}_{noise}$ is separated into two contributions, diagonal elements representing \textit{fluctuations} of the energy levels ($\tilde{H}_{fluc}$) and off-diagonal elements describing \textit{leakage} between dressed states ($\tilde{H}_{leak}$). After noting that $\partial_B\omega_{+1} + \partial_B\omega_{-1} = \partial_B\omega_{0'}$ (see \ref{app:dephasing}), one finds that fluctuations of the dressed state energy levels in (\ref{eq:Hfluc}) are all proportional to the clock sensitivity $\partial_B\omega_{0'}$. Therefore, despite being encoded with magnetic sensitive states, the dressed states are robust to dephasing and achieve coherence times on the order of the clock state $\ket{0'}$. The leakage Hamiltonian describes noise induced transitions among the dressed states. The first term dominates given that $\partial_B\omega_{+1} -\partial_B\omega_{-1} \gg \partial_B\omega_{0'} $ and the spins are sensitive to noise near the dressed splitting $\Omega_{dr}/\sqrt{2}$. Assuming that the magnetic field noise spectrum scales as $1/f^\alpha$, it is beneficial to increase $\Omega_{dr}$ so that the energy gap in the dressed state basis is larger and lower powers of noise are coupled.

\subsubsection*{Amplitude fluctuations} \label{sec:amplitude_fluctuations}\hspace*{\fill} \\

\indent Fluctuations of the microwave amplitude $\Omega_{dr}$ arise from imperfect microwave sources and noisy components such as temperature or current fluctuations in amplifiers. We consider all microwave fields to pass through the same microwave setup, hence common amplitude fluctuations can collectively be modelled with the replacement $\Omega_{dr} \rightarrow \Omega_{dr} + \delta\Omega(t)$. The state $\ket{D}$ is intrinsically insensitive to power fluctuations since its eigenenergy is independent of $\Omega_{dr}$. The $\ket{u}$ and $\ket{d}$ states however suffer dephasing from Rabi frequency fluctuations since their energies are proportional to $\Omega_{dr}/\sqrt{2}$. Despite this, the two-qubit states $\ket{du}$ and $\ket{ud}$ are degenerate and remain unaffected given that their eigenenergies are zero and independent of the dressing amplitude (as illustrated in figure (\ref{fig:energy_diagram})). 

Noise in the drive $\Omega_{dr}$ will still damage the fidelity through the first order expansion of the dressing Hamiltonian (equation (\ref{eq:Hdr1})) and the effects are two-fold. On the one hand, since the effective coupling strength $J_{eff}$ is dependent on $\Omega_{dr}$, amplitude fluctuations will lead to fluctuations of the interaction strength $J_{tot}$. These fluctuations, however, can usually be neglected in the limit $J_{eff} \ll J_0$. On the other hand, the dressing fields introduce errors due to off-resonant coupling to the spectator dressed states as well as coupling to the motional sidebands. Amplitude fluctuations may therefore vary the error term in equation (\ref{eq:eta_tot}). We expect however both of these error mechanisms to be negligible. 

\subsubsection*{Simulation Results}\hspace*{\fill} \\

The robustness of our scheme to magnetic and amplitude noise is demonstrated by means of numerical simulations and we compare it to existing methods. In Ref. \cite{piltz2016}, the J-coupling interaction is protected from dephasing by interleaving $\pi$ pulses throughout the evolution. This Pulsed Dynamical Decoupling (PDD) method refocuses $\sigma_z$ shifts and increases resilience to dephasing. In the same way that magnetic field noise near $\Omega_{dr}/\sqrt{2}$ is detrimental to the dressed states, the PDD scheme is affected by noise at frequencies $\omega/2\pi > 2N_\pi/\tau$ (the cutoff frequency may change with the timings of the pulses) \cite{biercuk2011}. The scheme is also subject to errors accumulated from imperfect $\pi$ pulses which stem from instrumental imperfections or amplitude and magnetic fluctuations.

\begin{figure}[t]
\centering
\includegraphics[scale=1]{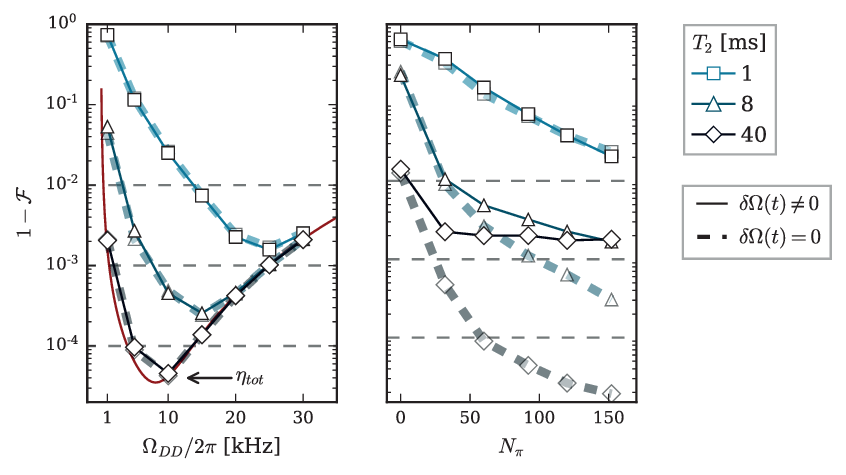} 
\caption{Infidelities under magnetic and amplitude fluctuations for continuous (this work, left) and pulsed (right) dynamical decoupling. Parameters are set to $\nu/2\pi = \SI{138.9}{kHz}$ and $\partial_z B = \SI{20}{T/m}$, resulting in $J_0/2\pi = \SI{40}{Hz}$ and gate durations $\SI{8.8}{ms}$ (left) and $\SI{6.3}{ms}$ (right) (without taking into account the $\pi$ pulse durations). The Rabi frequency of the $\pi$-pulses (right) is set to $\SI{50}{kHz}$ \cite{piltz2016}. Each point is averaged over $10^2$ realizations of two Ornstein-Uhlenbeck processes modelling magnetic and amplitude noise. We have assumed relative amplitude fluctuations of $\delta\Omega/\Omega = 5\times 10^{-3}$ and a correlation time of $\SI{0.5}{ms}$ \cite{cohen2016}. $T_2$ refers to the coherence time of the magnetic sensitive transition $\ket{0}\rightarrow\ket{+1}$. Data points connected by a dashed line are the result of simulations which only consider magnetic fluctuations and amplitude noise is set to zero. Solid lines consider both noise processes. The red line (left) plots the intrinsic error $\eta_{tot}$ of the continuous dynamical decoupling scheme.}
\label{fig:spin_decoherence}
\end{figure}

We show here that continuous dynamical decoupling (CDD) of the J-coupling gate outperforms the PDD scheme. For the former, the approximate Hamiltonian (\ref{eq:Happrox}) along with the noise Hamiltonian (\ref{eq:Hnoise}) are numerically integrated. The latter PDD scheme is simulated using the bare J-coupling Hamiltonian of equation (\ref{eq:HJint}). A XY4 sequence with $N_\pi$ $\pi$-pulses is implemented, where alternating rotations along the X and Y axes increases robustness to amplitude fluctuations \cite{viola1999}. Noise in the qubit frequency $\delta\omega_m(t)$ is modelled as an Ornstein-Uhlenbeck (OU) process (see Appendix B) \cite{wang1945}. Amplitude fluctuations $\delta\Omega(t)$ are also modeled as an OU process after replacing $\Omega_{dr} \rightarrow \Omega_{dr} + \delta\Omega(t)$ in all Hamiltonians. The diffusion constant in the OU model is chosen such that $c = 2(\Delta_\Omega \Omega_{dr})^2/\tau_c$ where $\Delta_\Omega = \delta\Omega(t)/\Omega_{dr}$ is the relative amplitude fluctuation and $\tau_c$ the correlation time \cite{lemmer2013}. The results are summarized in figure (\ref{fig:spin_decoherence}).

We first confirm that amplitude noise has virtually no effect on our gate scheme as numerical simulations with and without amplitude fluctuations yield the same fidelities. In contrast, amplitude noise has a large influence on the pulsed scheme. We observe a plateau for $T_2 = \SI{40}{ms}$ where the errors seemingly do not decrease despite adding more refocussing pulses. While increasing $N_\pi$ may increase the fidelity by mitigating dephasing of the bare states, the error accumulated from each refocussing $\pi$-pulse equally decreases the fidelity. In our simulations, errors in the $\pi$-pulses arise from magnetic and amplitude fluctuations only. Instrumental imperfections are not taken into account and we expect higher infidelities for larger $N_\pi$. As an example, in Ref. \cite{piltz2016} the number of pulses was optimized for a given evolution time and $N^{opt}_\pi = 60$ was found for $\tau = \SI{8.6}{ms}$. The simulation results of the PDD scheme can therefore be considered as a best case limit. 

From figure (\ref{fig:spin_decoherence}), we further observe that our CDD scheme results in higher fidelities for all $T_2$ times. Errors below the fault-tolerant threshold ($10^{-2}$) are possible for times $T_2 > \SI{1}{ms}$ and the smallest simulated infidelity is $\num{1.6e-3}$. With a coherence time of $\SI{8}{ms}$ ($\SI{40}{ms}$), the smallest infidelity is $\num{2.6e-4}$ ($\num{4.6e-5}$). For high coherence times, increasing the dressing power $\Omega_{dr}$ largely decreases infidelities due to dephasing, however the gate becomes limited by intrinsic error sources and the fidelity coincides with $\eta_{tot}$.

\subsection{Motional Decoherence}

Here we consider effects due to motional decoherence, namely \textit{thermal noise} and \textit{motional heating}. Thermal noise arises from the statistical nature of the ions' thermal states, whereas motional heating is due to coupling of the motion with a thermal bath which leads to changes of the thermal state over time. Frequency fluctuations of the motional modes which lead to motional dephasing are not considered as their effects are negligible. Changes of the motional frequencies will primarily lead to fluctuations of the J-coupling strength. Assuming a poor motional dephasing time of $\SI{10}{ms}$ and hence fluctuations of $\delta\nu = \SI{100}{Hz}$, the interaction strength is expected to vary by $\SI{0.04}{\%}$. From simulations, we verify that such changes have a negligible impact on the fidelity.

Entangling gates such as the M{\o}lmer-S{\o}rensen gate do not require ground state cooling of the motional modes as they are insensitive to the temperature $\bar{n}_{th}$ \cite{sorensen2000}. However, larger thermal states increase the gate's sensitivity to other sources of errors and ultimately limit fidelities. Initialization to the motional ground state is achieved with sideband cooling and has successfully been demonstrated with microwave radiation \cite{weidt2015}, although it may add an experimental overhead. Alternatively, the implementation of "hot" gates, i.e. gates that can achieve high fidelities despite large temperatures, may offer several significant advantages. For example, the gate would become tolerant to operations within QCCD (quantum charge-coupled device) proposals \cite{kielpinski2002, pino2021} which vary the temperature (e.g. shuttling). Furthermore, the complexity of sympathetic cooling in scalable architectures would also be simplified as sideband cooling is no longer required.

Beyond the dependence on the temperature, entangling gate fidelities are often limited by heating of the vibrational modes. The heating rate depends on the electric field noise, therefore considerable effort has been devoted to improving the quality of microfabricated surface ion traps and designing low noise electronics. In some cases, cryogenic cooling is employed to lower the temperature of the chip \cite{romaszko2020}. All these solutions however pose an important engineering challenge and add a considerable experimental overhead. 

\begin{figure}[t]
\centering
\includegraphics[scale=.9]{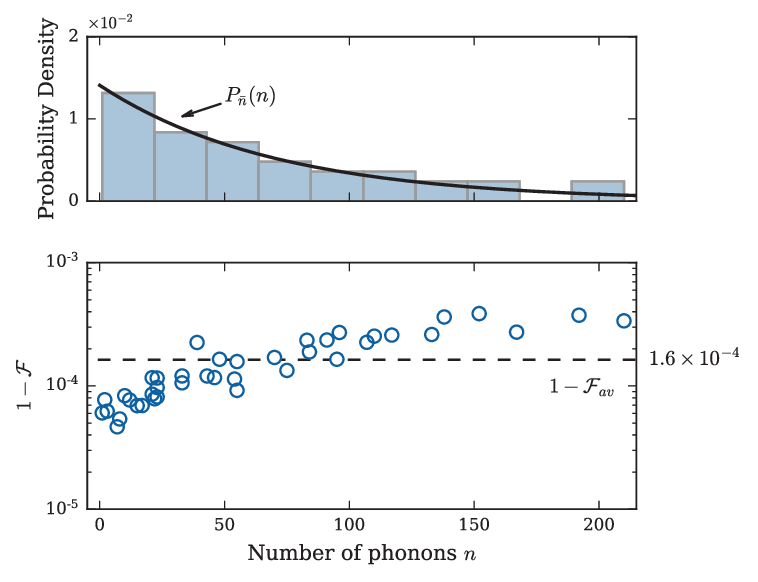} 
\caption{(Top) Probability density of a thermal state with temperature $\nbar = 70$. Solid line is the theoretical distribution $P_{\nbar}(n)$. Bars represent the histogram of 40 random variables sampled from $P_{\nbar}(n)$ and correspond to the data points of the bottom figure. (Bottom) Infidelities from motional decoherence obtained by numerically simulating $H_{exact}$ using the MCWF method. Each point is the result of 20 trajectories. The secular frequency is $\nu_1/2\pi = \SI{140}{kHz}$, the gradient $\partial_zB = \SI{20}{T/m}$, the Rabi frequency is set to $\Omega_{dr}/2\pi = \SI{8}{kHz}$ and the heating rate is $\ndot = \SI{100}{s^{-1}}$. The average fidelity is indicated by the dashed line. }
\label{fig:motional_decoherence}
\end{figure}

The intrinsic J-coupling interaction is robust to motional decoherence provided that the ions remain in a harmonic potential \cite{khromova2012, piltz2016}. We note that a similar robustness is possible with, for example, a far-detuned M{\o}lmer-S{\o}rensen gate which, in the weak-field regime, leaves the spins disentangled from the motion throughout the evolution \cite{sorensen2000}.  This remains, however, an approximation in the limit of large detunings\footnote{It is also interesting to consider the robustness that is required from a multi-loop MS gate to obtain similar fidelities to that of a dressed J-coupling gate. The requirements are roughly approximated from the infidelity model of \cite{sorensen2000}. For a $\SI{608}{\mu s}$ duration (assuming the same parameters as figure (\ref{fig:motional_decoherence}) and $\Omega_{MS} = \SI{40}{kHz}$), the resulting infidelity due to motional heating is $2\times 10^{-2}$. In order to obtain errors near $10^{-4}$, one would have to perform approximately $\num{40000}$ loops, increasing the gate duration to $\SI{122}{ms}$. This is much larger than what could be achieved with the J-coupling gate, for which the duration is $\SI{8.86}{ms}$ (c.f. figure (\ref{fig:motional_decoherence})).}. Furthermore, this differs from the robustness of the intrinsic J-coupling interaction which does not entangle the spin and motion throughout the evolution. In our proposal, the introduction of continuous drives adds an off-resonant interaction with the motional sidebands which was shown to limit fidelities for high dressing powers (section (\ref{sec:calculating_neglected_terms})). Motional decoherence may therefore further increase errors due to this coupling. The leading terms of the sideband Hamiltonian (\ref{eq:Hdr1_ME}) were found, however, to not include any motional operators and therefore remain insensitive to motional decoherence. In what follows, we must therefore consider the exact Hamiltonian.

In the absence of sideband cooling, the smallest achievable temperature of the $n^{th}$ vibrational mode is determined by the Doppler cooling limit $\nbar^{(n)}_{min} = \Gamma/(2\nu_n)$ where $\Gamma$ is the natural linewidth of the cooling transition. In \yb, the natural linewidth of the $^2$S$_{1/2},F=1 \rightarrow ^2$P$_{1/2}, F=0$ cooling transition is $\Gamma/2\pi = \SI{19.6}{MHz}$ \cite{olmschenk2005}. Since smaller vibrational frequencies are favourable for faster gates, the initial temperature is large and for $\nu_1/2\pi = \SI{140}{kHz}$, the Doppler cooling limits for $N=2$ ions are $\nbar^{(1)}_{min} = 70$ and $\nbar^{(2)}_{min} = 40.4$. Despite these large temperatures, we remain in the Lamb-Dicke regime $\epsilon \sqrt{n} \ll 1$. 

Numerical simulations of large thermal states are computationally intensive as the number of elements in the density matrices scales as $M\times M$, where $M$ is the dimension of the vibrational Hilbert space. For two ions and therefore two axial modes of vibration, the computational complexity becomes $\mathcal{O}(M^4)$ and is infeasible for large temperatures such as those set by the Doppler cooling limit. In order to reduce the complexity, we first truncate the first order sideband Hamiltonian of equation (\ref{eq:Hdr1}) to only include the center-of-mass (COM) mode and neglect the stretch mode which reduces the complexity to $\mathcal{O}(M^2)$. This choice is justified by : (i) the heating rates associated with vibrational modes of differential motion (such as the out-of-phase motion of the stretch mode) are orders of magnitude smaller (ii) the Doppler cooling limited temperatures of higher modes are smaller hence thermal noise from the COM mode will dominate (iii) the frequency of the stretch mode is further from the carrier and therefore off-resonant coupling to the motional sideband is smaller. The computational complexity is further reduced by employing a Monte-Carlo wave-function (MCWF) approach \cite{molmer1993}. Instead of solving a master-equation with a density matrix of size $M^2$, the MCWF method stochastically propagates a pure state given a non-hermitian Hamiltonian over multiple averaged trajectories. By employing pure states, the number of elements is reduced to $M$. Since the thermal state $\rho_{\nbar}$ under consideration is a statistical mixture of pure states where $\rho_{\nbar} = \sum_n^M P_{\nbar}(n)\ket{n}\bra{n}$ and $P_{\nbar}(n) = \nbar^{n}/(1 + \nbar)^{(n+1)}$ is the Bose-Einstein probability distribution, pure Fock states $\ket{n}$ can be randomly sampled from $P_{\nbar}(n)$ and evolved following the MCWF approach.

The results of the numerical simulations are presented in figure (\ref{fig:motional_decoherence}). The exact Hamiltonian of equation (\ref{eq:Hexact}) is considered and motional heating is modelled by introducing the operators $C_1 = \sqrt{\ndot \nbar}a^\dag$ and $C_2 = \sqrt{\ndot (1+\nbar)}a$, where a heating rate of $\ndot = \SI{100}{s^{-1}}$ was considered. Initial Fock states $\ket{n}$ are sampled from the discrete probability distribution $P_{\nbar}(n)$ where the temperature $\nbar = 70$ is set by the Doppler cooling limit. The maximum allowable Fock state is set to $n_{max} = 250$ which contains $95\%$ of the distribution. A sample size of $S = 40$ was chosen and shows good agreement with the ideal probability distribution $P_{\nbar}(n)$. Each initial Fock state is propagated with the MCWF approach over $20$ trajectories and the averaged fidelity results in one data point in figure (\ref{fig:motional_decoherence})\footnote{The longest execution time was 10 hours (Intel Xeon W-2102).}. As expected, the infidelity increases for higher Fock states and the largest simulated error was $3.9\times 10^{-4}$. The total fidelity is obtained by averaging over all samples and we find $1 - \mathcal{F}_{av} = 1.6 \times 10^{-4}$ which is well below the fault-tolerant threshold. The infidelity in the absence of motional decoherence due to intrinsic errors is $1 - \mathcal{F}_0 = 1.8 \times 10^{-5}$. Motional decoherence therefore only modestly contributes to the overall error and we do not expect larger heating rates $> \SI{100}{s^{-1}}$ to significantly impact the fidelity. 

%% file: sections/section4.tex
\section{Increasing the gate speed} \label{sec:fast_gates}

In order to obtain faster gates, one must increase the J-coupling by means of introducing a higher magnetic gradient or decreasing the secular frequency. In section (\ref{sec:calculating_neglected_terms}) we identified the dominant sources of errors and showed that high J-coupling strengths may lead to considerable infidelities due to off-resonant coupling to spectator dressed states. Therefore, although the secular frequency and gradient may be tuned to decrease the gate duration, the infidelity may become appreciable.

\begin{figure}[t]
\centering
         \includegraphics[scale = 1]{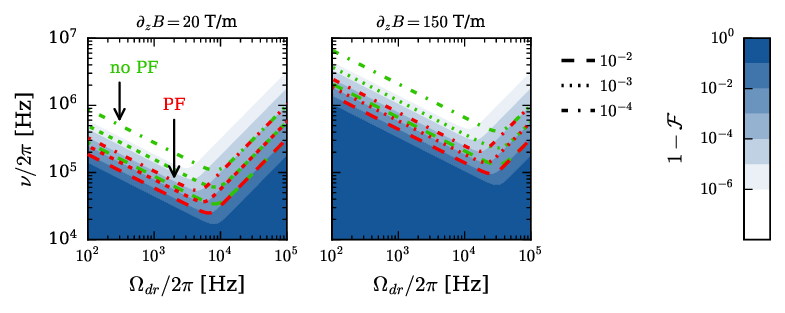}
        \caption{Infidelities after implementing a phase flip (PF) at half the gate duration. The modified error function is obtained from equation (\ref{eq:eta2_pf}). The infidelities without a phase flip (no PF) are plotted for reference (red) and follow the same legend. Dashed, dotted and dash-dotted lines represent infidelity levels at $10^{-2}$, $10^{-3}$ and $10^{-4}$.}
        \label{fig:phase_flip}
\end{figure}

Infidelities introduced by higher J-coupling strengths can be mitigated by introducing a phase flip at half the gate duration such that $\Omega_{dr} \rightarrow - \Omega_{dr}$. This will refocus unwanted oscillations which leak population out of the computational subspace through off-resonant coupling to the spectator dressed states. In order to see the improved performance from a phase flip, we derive the corresponding evolution operator $U = e^{-i\frac{\tau}{2}\tilde{H}'}e^{-i\frac{\tau}{2}\tilde{H}}$ where $\tilde{H}$ is the Hamiltonian of equation (\ref{eq:Hj_time_ind}) containing the gate operator and off-resonant coupling terms, and $\tilde{H}'(t)$ is the modified Hamiltonian with a phase flip. The new amplitude $a'$ of the state $\ket{DD}$ previously derived in equation (\ref{eq:amp_dd_state}) is

\begin{align}
& a'_{DD} = a_{DD} + \frac{2J_0^2\Omega_{dr}^2}{4J_0^2 + \Omega_{dr}^4} (\beta_1' - \beta_2')^2, \\
& \beta_1' = \cos{(\frac{\pi\sqrt{2J_0^2 + \Omega_{dr}^2 + \sqrt{4J_0^4 + \Omega_{dr}^4}}}{2\sqrt{2}J_0})},\\
& \beta_2' = \cos{(\frac{\pi\sqrt{2J_0^2 + \Omega_{dr}^2 - \sqrt{4J_0^4 + \Omega_{dr}^4}}}{2\sqrt{2}J_0})}.
\end{align}

The presence of a phase flip introduces an additional higher order term in the $\ket{DD}$ state's amplitude. A new upper bound on the infidelity is found by setting $\beta_1'\rightarrow 1$ and equation (\ref{eq:eta_2}) becomes 

\begin{equation}\label{eq:eta2_pf}
\eta_2 = 4\frac{J_0^4}{\Omega_{dr}^4}.
\end{equation}

Adding a phase flip at half the gate duration therefore leads to a quadratic improvement of the fidelity. The new infidelity term $\eta_{tot}$ is plotted in figure (\ref{fig:phase_flip}). One can see that the infidelities at lower dressing powers $\Omega_{dr}$ are much smaller for the same secular frequencies, which ultimately allows for faster entangling gates. We recall that for a gradient of $\SI{150}{T/m}$, the smallest gate duration to achieve errors $< 10^{-4}$ is $\tau = \SI{1.39}{ms}$. If a phase flip is introduced, the gate duration can be reduced to $\tau = \SI{345}{\mu s}$ ($\nu_1/2\pi = \SI{205.8}{kHz}$, $\Omega_{dr}^{opt}/2\pi = \SI{17.2}{kHz}$).

%% file: sections/section5.tex
\section{Applications and architectures} \label{sec:applications_and_architectures}

In this section, we outline various applications and advantages of the continuously dressed J-coupling gate as well as its integration with existing trapped ion QIP architectures. We also discuss a potential shortcoming due to voltage noise on the electrodes.

We first consider the mode of operation of our gate scheme within a quantum circuit. Considering a string of ions, an entangling gate can be achieved between any pair of ions in the chain, although the J-coupling interaction strength will vary and is usually much weaker for long-range couplings. For a small number of qubits, a pair of MW dressing fields could be used for each ion such that the dressed state $\ket{D}$ is always active and all computations within a circuit are carried out in the $\{ \ket{0'}, \ket{D}\}$ basis. High fidelity quantum control has been demonstrated with RF radiation in this basis \cite{randall2015}. A unique Rabi frequency would be associated to each ion, and an entangling gate would involve instantaneously changing the microwave powers of two qubits such that they're equal. This solution however scales poorly for longer chains as each ion requires a pair of microwave fields, whether it is idle of participating in a gate. Alternatively, one could use the bare $\{\ket{0}, \ket{0'}\}$ basis for memory and single qubit rotations as proposed in Ref. \cite{piltz2016} while the dressed state basis is used solely for two qubit gates. High fidelity mapping between both bases was demonstrated in Ref. \cite{randall2018}. The number of MW fields now scales with the number of different types of gates and allows for resource-efficient quantum circuits.

An advantage of our scheme is the ease with which one can obtain parallel entangling gates within a long ion chain, which allows for better algorithmic scalability. Since geometric phase gates use the common modes of vibration as an information bus, parallel operations within an ion chain may excite the same modes and impact the gate's fidelity. Various solutions have been proposed \cite{bentley2020,grzesiak2020}, however these add an experimental overhead. Our scheme could trivially be extended to allow the simultaneous execution of any number of two qubit entangling operations by choosing a unique Rabi frequency for each of the ion pair's dressing fields. Assuming that the $\{\ket{0}, \ket{0'}\}$ basis is used for memory and single qubit gates, the number of microwave fields scales with the number of parallel entangling gates. Note that the bare J-coupling interaction does not allow for parallel gates as there are no means of controlling the spin-spin coupling. This parallelism is possible due to the unique controllability of the J-coupling interaction by virtue of the continuous drives.

We now discuss a potential shortcoming of our scheme due to hardware implementations. The J-coupling interaction strength is inversely proportional to the secular frequency, therefore lowering $\nu$ may be desired to increase the gate speed. Lowering the secular frequency however involves decreasing the depth of the axial trapping potential which makes the ions' displacements more prone to external perturbations. For ions in a magnetic gradient, noise of the position directly transforms into magnetic field noise which decoheres the spin states. A potentially dominant source of noise is voltage fluctuations on the electrodes which could result from noisy instrumentation. One finds that the conversion of voltage noise to magnetic field noise is proportional to $\propto \partial_zB^2/\nu_z^4$ (see \ref{app:voltage_noise}). Recalling that $\tau \propto \nu^2/\partial_zB^2$, one can see that while decreasing the secular frequency may increase the gate speed, the noise will worsen at a faster rate. Therefore, reducing the secular frequency may be detrimental to the entangling gate. Similarly, increasing the magnetic field gradient will increase both the gate speed and the noise at the same rate and can therefore not be used as a means to improve the fidelity. However, it may still be beneficial to increase the gradient in order to lower the interaction time. Note that many ion traps such as the one in Ref. \cite{gulde2003} are intrinsically robust to voltage noise. In this configuration, the trapping DC electrodes are symmetric about the ion chain's position and parasitic forces due to noise at opposing ends cancel each other out.

Alternatively, microfabricated surface traps may be preferred due to their scalability. One such proposed architecture makes use of global microwave radiation fields and functional zones across an array of X-junctions, and shuttling provides all-to-all connectivity  \cite{lekitsch2017}. Our gate scheme is compatible with this architecture as only a handful of microwave fields are required to address any number of resonant ions in a suitable gate zone. Our gate scheme further eliminates an important engineering challenge associated with global microwave radiation. Entangling gates whose interaction strength is proportional to the field's amplitude require the spatial variation of the microwave amplitude to be minimized across the chip to obtain identical evolution times in every gate zone. Alternatively, different gate durations can be used in each zone to account for amplitude variations. However, this adds some experimental overhead. The constraint of amplitude homogeneity disappears with the use of dressed J-coupling gates since the interaction strength is only dependent on the secular frequency and the gradient, both of which are tunable in each individual gate zone by local control of current and voltages. Although the dressing Rabi frequency $\Omega_{dr}$ may vary across the chip, one can fairly assume that the variation between two ions within a gate zone is negligible and therefore the fidelity is not compromised. Note that these considerations are true as long as the amplitude variations of the global fields are the same for every frequency tone.

%% file: sections/conclusion.tex
\section{Conclusion}

We have proposed an entangling gate which uses the intrinsic J-coupling of ions in a static magnetic field gradient. Dephasing of the magnetic sensitive states is suppressed by introducing continuous microwave dressing fields which lead to a clock-like protected subspace, in which a pseudo spin-spin interaction takes place. The continuous fields open up a new energy gap that decouple the qubit from low frequency noise. The interaction is virtually insensitive to amplitude fluctuations of the dressing fields, as all two-qubit states populated during the gate's evolution have energy levels that are decoupled from the dressing fields' power. We have shown that our scheme outperforms existing pulsed dynamical decoupling solutions which remain sensitive to pulse imperfections. Furthermore, numerical simulations predict that infidelities well below the fault-tolerant threshold are possible for reasonable magnetic field gradients ($\SI{20}{T/m}$) despite coherence times of the bare states $\approx \SI{1}{ms}$. Fault-tolerance is therefore within reach of ready-available quantum hardware, while longer coherence times have already been demonstrated with improvements such as magnetic shielding \cite{wang2021}. A better performance is expected from larger magnetic field gradients that would increase the gate speed. We further show that adding a phase flip on the dressing fields increases the resilience of the gate to intrinsic error mechanisms and allows for faster gates. Despite the inherently slow nature of the J-coupling interaction, reasonable gate durations are achievable with realistic experimental parameters.

The gate is virtually insensitive to motional decoherence, as numerical simulations suggest that high fidelities ($\approx 10^{-4}$) are achievable despite large temperatures ($\nbar = 70$) and in the presence of motional heating ($\ndot = \SI{100}{s^{-1}}$). This robustness reduces the experimental overhead associated with scalable QCCD architectures as the gate is insensitive to operations such as shuttling which vary the temperature. The complexity of sympathetic cooling is also alleviated as sideband cooling is not required. The robustness to motional heating could further eliminate the need for cryogenic cooling.

Our scheme is scalable both in terms of physical and algorithmic resources. Ions that are not participating in an entangling sequence can be mapped to a memory clock state which does not require continuous driving. Furthermore, the use of dressing fields introduces a previously unachievable controllability and selectivity of the J-coupling interaction which allows for parallel execution of two-qubit gates. The number of required fields then scales with the amount of simultaneous entangling gates. We finally show that our entangling scheme is compatible with existing scalable architectures that use global microwave fields. The robustness to amplitude fluctuations removes the stringent requirement of homogeneity of the global field's amplitude across large chips. 

The dressed J-coupling gate proposed in this work can be useful both in near- and long-term quantum hardware. Parallelism and controllability allow resource efficient quantum circuits to run on near-term traps. Fast deployment of quantum processing units could be possible thanks to the robustness of this scheme and its minimal hardware requirements. In the long-term, difficult tasks such as cryogenic cooling and sympathetic sideband cooling could become redundant. 

%% file: sections/acknowledgements.tex
\section{Acknowledgements}

We gratefully acknowledge helpful discussions with Alex Retzker and Daniel Cohen. This work was supported by the U.K. Engineering and Physical Sciences Research Council via the EPSRC Hub in Quantum Computing and Simulation (EP/T001062/1), the U.K. Quantum Technology hub for Networked Quantum Information Technologies (No. EP/M013243/1), the European Commission’s Horizon-2020 Flagship on Quantum Technologies Project No. 820314 (MicroQC), the U.S. Army Research Office under Contract No. W911NF-14-2-0106, the Office of Naval Research under Agreement No. N62909-19-1-2116, the University of Sussex, and through a studentship in the Quantum Systems Engineering Skills \& Training Hub at Imperial College London funded by the EPSRC (EP/P510257/1).

%% file: sections/appendix1.tex
\section{Magnus Expansion of the first order Sideband}
\label{app:Magnus_expansion}

In section (\ref{sec:calculating_neglected_terms}) of the main text, the dominant terms of the first order Hamiltonian $H^1_{dr}$ were used to derive an approximate error term arising from the off-resonant interaction of the dressing microwave fields with the motional sidebands. Here, we present detailed steps of the Magnus Expansion (ME) which resulted in equation (\ref{eq:Hdr1_ME}). We closely follow the derivation and notation used in Ref. \cite{lemmer2013} as it is clear and pertinent to our Hamiltonian.

Let us recall the first order Hamiltonian (\ref{eq:Hdr1}),

\begin{equation} \label{eq:Hdr1_appendix}
\tilde{H}_{dr}^1 = - \sum_{j}\sum_n \frac{\hbar\epsilon_{j, n} \Omega_{dr}}{2} (\tilde{a}_n^\dagger e^{ i \nu_n t} - \tilde{a}_n e^{- i \nu_n t})(S_+^{(j)} e^{i \frac{\Omega_{dr}}{\sqrt{2}}t} - S_-^{(j)} e^{- i \frac{\Omega_{dr}}{\sqrt{2}}t}).
\end{equation}

\indent The first and second order terms of the Magnus Expansion (ME) are
\begin{align}
& \Omega_1(t, t_0) = - \frac{i}{\hbar} \int_{t_0}^t dt_1H(t_1), \label{eq:ME_Om1} \\
& \Omega_2(t, t_0) = - \frac{1}{2\hbar^2}\int_{t_0}^t dt_1\int_{t_0}^{t_1}dt_2 [H(t_1), H(t_2)]. \label{eq:ME_Om2}
\end{align}

\indent Higher order terms are neglected in the Lamb-Dicke regime. The first order expansion of $\tilde{H}_{dr}^1$ (\ref{eq:Hdr1_appendix}) according to (\ref{eq:ME_Om1}) is

\begin{align} \label{eq:Om1}
\Omega_1(t,0) = \sum_n \sum_{j}& \frac{\epsilon_{j,n}\Omega_{dr}^2}{2(2\nu_n^2 - \Omega_{dr}^2)}(\tilde{a}_n^\dagger(e^{-it(\nu_n + \frac{\Omega_{dr}}{\sqrt{2}})} - 1) - \tilde{a}_n(e^{it(\nu_n - \frac{\Omega_{dr}}{\sqrt{2}})} - 1))(S_-^{(j)} + S_+^{(j)}) \nonumber \\ &- \frac{\epsilon_{j,n}\nu_n\Omega_{dr}}{\sqrt{2}(2\nu_n^2 - \Omega_{dr}^2)}(\tilde{a}_n^\dagger(e^{-it(\nu_n + \frac{\Omega_{dr}}{\sqrt{2}})} - 1) + \tilde{a}_n(e^{it(\nu_n - \frac{\Omega_{dr}}{\sqrt{2}})} - 1))(S_-^{(j)}- S_+^{(j)}).
\end{align}

This expression describes an effective spin and motion coupling within the dressed basis and leads to sideband transitions. The second order expansion is separated into three parts, such that 
\begin{equation} \label{eq:Om2}
\Omega_2(t,0) =\Omega_2^a(t,0) + \Omega_2^b(t,0) + \Omega_2^c(t,0).
\end{equation}

The first expression $\Omega_2^a(t,0)$ corresponds to terms which are time-independent :

\begin{align} \label{eq:Om2a}
\Omega_2^a(t, 0) = \sum_n & 2\mathcal{F}_{j=1,k=2,n}^1(S_+^{(1)}S_+^{(2)} - S_-^{(1)}S_-^{(2))})  \nonumber \\ & + \sum_{j} \mathcal{F}_{jjn}^1(S_+^{(j)}S_+^{(j)} - S_-^{(j)}S_-^{(j)})  + \mathcal{F}^2_{jn}((\tilde{a}^\dagger)^2 - \tilde{a}^2)S_z^{(j)}.
\end{align}

We recall that $S_z^{(j)} = \ket{u}\bra{u} - \ket{d}\bra{d}$. The coupling strengths are
\begin{equation}
\begin{aligned}
& \mathcal{F}^1_{jkn} = \frac{\epsilon_{j,n}\epsilon_{k,n}\Omega_{dr}\nu_n}{2\sqrt{2}(2\nu_n^2 - \Omega_{dr}^2)}, \\
& \mathcal{F}^2_{jn} = \frac{\epsilon_{j,n}^2\Omega_{dr}^3}{4\sqrt{2}\nu(2\nu_n^2 - \Omega_{dr}^2)}.
\end{aligned}
\end{equation}

The first two terms of equation (\ref{eq:Om2a}) describe single spin operations. The last term is another effective spin-motion coupling. The second expression $\Omega^b_2(t,0)$ collects terms which increase linearly in time, 

\begin{align} \label{eq:Om2b}
\Omega_2^b(t,0) = i t \sum_n  \Big[ & 2\tilde{\mathcal{G}}^1_{1,2, n} \left( S_+^{(1)}S_-^{(2)}  + S_-^{(1)}S_+^{(2)}\right) \nonumber \\ 
& + \sum_j \tilde{\mathcal{G}}_{j,j, n}^1 \left( S_+^{(j)}S_-^{(j)} + S_-^{(j)}S_+^{(j)} \right)  + \tilde{\mathcal{G}}^2_{j, n}  \tilde{a}^\dagger\tilde{a}S_z^{(j)} \Big],
\end{align}

with coupling constants

\begin{equation}
\begin{aligned}
& \tilde{\mathcal{G}}_{jkn}^1 = \frac{\epsilon_{j,n}\epsilon_{k,n}\Omega_{dr}^2 \nu_n}{2(2\nu_n^2 - \Omega_{dr}^2)},  \\
& \tilde{\mathcal{G}}_{jn}^2 = \frac{\epsilon_{j,n}^2\Omega_{dr}^3 }{\sqrt{2}(2\nu_n^2 - \Omega_{dr}^2)} .
\end{aligned}
\end{equation}

The first term of $\Omega^b_2(t,0)$ (\ref{eq:Om2b}) is another J-coupling interaction identical to equation (\ref{eq:Hg}) of the main text. The second term describes single qubit spin operations. The last term describes a shift of the $\ket{u}$ and $\ket{d}$ states proportional to the motional temperature.

The final expression $\Omega^c_2(t,0)$ of equation (\ref{eq:Om2}) consists of rotating terms, 

\begin{align}
\Omega_2^c(t,0) = \sum_n & \sum_{j}(\mathcal{M}^1_{jjn}-\mathcal{M}^2_{jjn}) \tilde{a}^\dagger_n\tilde{a}_n S_z^{(j)} - (\mathcal{M}^3_{jn}(\tilde{a}^\dagger_n)^2 - \mathcal{M}^{3*}_{jn} \tilde{a}_n^2)S_z^{(j)} \nonumber \\ & - (\mathcal{M}^1_n\hat{c}^1 + \mathcal{M}^2_n\hat{c}^{2} )
 + \mathcal{M}^4_n\hat{c}^3 - \mathcal{M}^{4*}_n\hat{c}^{3*}.
\end{align}

The coupling constants are

\begin{align}
& \mathcal{M}^1_{jkn} = \frac{\epsilon_{j,n}\epsilon_{k,n}\Omega_{dr}^2}{8(\nu+ \frac{\Omega_{dr}}{\sqrt{2}})^2}(e^{it(\nu+\frac{\Omega_{dr}}{\sqrt{2}})} - e^{-it(\nu + \frac{\Omega_{dr}}{\sqrt{2}})}), \\
& \mathcal{M}^2_{jkn} = \frac{\epsilon_{j,n}\epsilon_{k,n}\Omega_{dr}^2}{8(\nu - \frac{\Omega_{dr}}{\sqrt{2}})^2}(e^{it(\nu-\frac{\Omega_{dr}}{\sqrt{2}})} - e^{-it(\nu - \frac{\Omega_{dr}}{\sqrt{2}})}) , \\
&  \mathcal{M}^3_{jn} = \frac{\epsilon_{j,n}\epsilon_{k,n}\Omega_{dr}^2}{4(2\nu^2 - \Omega_{dr}^2)}(e^{it(\nu - \frac{\Omega_{dr}}{\sqrt{2}})} - e^{it(\nu + \frac{\Omega_{dr}}{\sqrt{2}})}) + \frac{\epsilon_{j,n}\epsilon_{k,n}\Omega_{dr}^3}{4\sqrt{2}\nu(2\nu^2 - \Omega_{dr}^2)}e^{-2it\nu}, \\
&  \mathcal{M}^4_{jkn} = \frac{\epsilon_{j,n}\epsilon_{k,n}\Omega_{dr}^2}{4(2\nu^2 - \Omega_{dr}^2)}(e^{it(\nu + \frac{\Omega_{dr}}{\sqrt{2}})} - e^{-it(\nu - \frac{\Omega_{dr}}{\sqrt{2}})}) - \frac{\epsilon_{j,n}\epsilon_{k,n}\nu\Omega_{dr}}{2\sqrt{2}\nu(2\nu^2 - \Omega_{dr}^2)}e^{\sqrt{2}it\Omega_{dr}}.
\end{align}

and spin operators 

\begin{align*}
& \hat{c}^1 = (S_-^{(1)} + S_-^{(2)})(S_+^{(1)} + S_+^{(2)}), \\
& \hat{c}^2 = (S_+^{(1)} + S_+^{(2)})(S_-^{(1)} + S_-^{(2)}), \\
& \hat{c}^3 = (S_+^{(1)} + S_+^{(1)})^2.
\end{align*}

The terms of $\Omega^c_2(t,0)$ are similar to those in (\ref{eq:Om2a}) however they rotate at frequencies proportional to $\nu$ and $\Omega$. 

The first order expansion $\Omega_1(t)$ and second order rotating terms $\Omega_2^c(t)$ are neglected under a RWA in the limit $\Omega_{dr} \ll \nu$. The time independent terms of $\Omega_2^a(t)$ are also dropped as they have no physical effect. The only terms which are left are the linear terms belonging to $\Omega^b_2(t,0)$, hence $\Omega(t,0) \approx \Omega_2^b(t,0)$.

After recalling that $U_{eff}(t) = e^{\Omega(t,0)}=e^{- \frac{it}{\hbar}H_{eff}}$, an effective Hamiltonian is found,
\begin{align}
H_{eff} = \ - \hbar \sum_n \Big[ & 2\tilde{\mathcal{G}}^1_{1,2, n} \left( S_+^{(1)}S_-^{(2)}  + S_-^{(1)}S_+^{(2)} \right) \nonumber \\ 
& +  \sum_j \tilde{\mathcal{G}}_{j,j, n}^1 \left( S_+^{(j)}S_-^{(j)} + S_-^{(j)}S_+^{(j)} \right)  + \tilde{\mathcal{G}}^2_{j, n}  \tilde{a}^\dagger\tilde{a}S_z^{(j)} \Big]
\end{align}

 This final expression corresponds to equation (\ref{eq:Hdr1_ME}) of the main text. 

%% file: sections/appendix2.tex
\section{Numerical Simulations of Dephasing}
\label{app:dephasing}

Fluctuations of the magnetic field $\delta B(t)$ at the position of the ion lead to fluctuations of their respective energy levels $\delta \omega_m(t)$ such that

\begin{equation}
\delta\omega_m(t) = \partial_B\omega_m \delta B(t),
\end{equation}

where the sensitivities $\partial_B\omega_m$ can be found from the Breit-Rabi formula \cite{breit1931}. For the $m^{th}$ state of the $F=1$ triplet of ytterbium,

\begin{align}
&\partial_B\omega_{+1} = \frac{\omega_0}{2}(\xi + \frac{\xi^2B}{\sqrt{1 + (\xi B)^2}}), \label{eq:sensitivity_p1}\\
&\partial_B\omega_{-1} = \frac{\omega_0}{2}(-\xi + \frac{\xi^2B}{\sqrt{1 + (\xi B)^2}}), \label{eq:sensitivity_m1}\\
&\partial_B\omega_{0'} = \omega_0(\frac{\xi^2B}{\sqrt{1 + (\xi B)^2}}).
\end{align}

Here we've defined $\xi = \frac{g_J \mu_B}{\hbar \omega_0}$, where $g_J$ is the electronic g-factor, $\mu_B$ the Bohr magneton and $\omega_0$ the unperturbed transition frequency. The sensitivities are assumed constant around a magnetic field $B_0$ and equal for all ions in a chain. From these derived sensitivities, we find that $\partial_B\omega_{+1} + \partial_B\omega_{-1} = \partial_B\omega_{0'}$. Furthermore, the robustness of the clock state is verified by noting that $\partial_B\omega_{\pm1} \gg \partial_B\omega_{0'}$.

\begin{figure}[t]
\centering
\includegraphics[scale=.7]{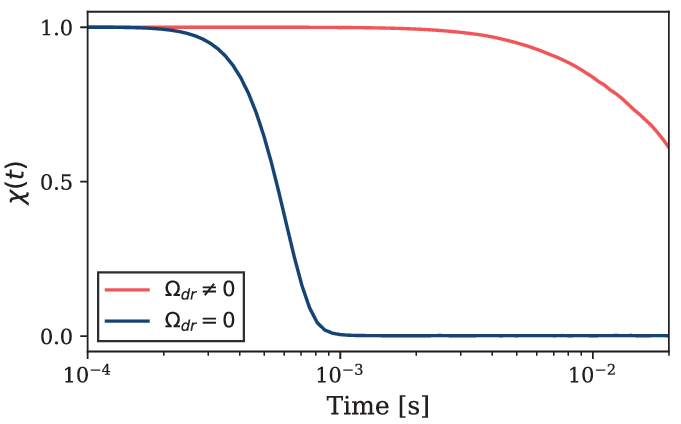} 
\caption{Decay due to dephasing with and without continuous drivings. The decay function is $\chi(t) = e^{-\langle \Delta\phi^2\rangle/2}$. The coherence time is $T_2^{(+1)} = 0.5$ ms, which fixes the OU parameters $\tau_c$ and $c$. Decay curves are averaged over $10^3$ realizations of the OU process. }
\label{fig:dephasing_comparison}
\end{figure}

Dephasing is described by the following noise Hamiltonian,

\begin{align} \label{eq:Hnoise_app}
H_{noise} = \sum_j \sum_{m} \frac{ \delta\omega_m(t)}{2}  \sigma_z^{(j,m)},
\end{align}

where the stochastic variable $\delta\omega_m(t)$ is an OU process with zero mean $\textrm{E}[\delta\omega (t)] = 0$ and variance $\textrm{Var}[\delta\omega(t)] = \frac{c\tau_c}{2}(1 - e^{-2t/\tau_c})$ \cite{wang1945, lemmer2013}. The parameters of the OU process are fixed by the coherence time $T_2$. The correlation time $\tau_c$ must satisfy $\tau_c \ll T_2$, and we choose $\tau_c = T_2/100$. The diffusion constant $c$ is then calculated as $2/T_2 \tau_c^2$.

We define the OU process with respect to the coherence of the $\ket{+1}$ state $T_2^{(+1)}$. In this way, the stocastic variables of (\ref{eq:Hnoise_app}) are all proportional to $\delta\omega_{+1}(t)$ such that $\delta\omega_{-1}(t) = \frac{\partial_B\omega_{-1}}{\partial_B\omega_{+1}}\delta\omega_{+1}(t)$ and  $\delta\omega_{0'}(t) = \frac{\partial_B\omega_{0'}}{\partial_B\omega_{+1}}\delta\omega_{+1}(t)$. To validate the choice of the OU process, dephasing during free induction decay is simulated with $H_{noise}$ (figure (\ref{fig:dephasing_comparison})). Dephasing under continuous driving of the dressing fields is also simulated by adding the dressing Hamiltonian $\tilde{H}^0_{dr}$ (equation (\ref{eq:Hdr0})). The coherence time is increased by more than an order of magnitude under continuous driving. 

When considering multiple ions in a static magnetic gradient, we recall that the absolute magnetic field at each ion is different. The sensitivities of the $\ket{\pm 1}$ states remain approximately constant given that the first term of equations (\ref{eq:sensitivity_p1}) and (\ref{eq:sensitivity_m1}) dominate ($\partial_B\omega_{\pm 1} \approx \pm \frac{\omega_0\xi}{2}$). The sensitivity of the clock transition, however, is more susceptible to change for different magnetic fields. We remedy this by evaluating the sensitivities at the ions' average magnetic field. In the simulation results of the main text (section (\ref{sec:robustness_noise})), an average magnetic field of $B_0 = \SI{7.5}{G}$ is considered with leads to $\frac{\partial_B\omega_{-1}}{\partial_B\omega_{+1}} = \num{-0.9967}$ and $\frac{\partial_B\omega_{0'}}{\partial_B\omega_{+1}} = \num{3.316e-3}$.

%% file: sections/appendix3.tex
\section{Voltage noise on the electrodes}
\label{app:voltage_noise}

A string of trapped ions is confined by a pseudopotential created by a combination of static and oscillating voltages on DC and RF electrodes. Any voltage noise on the electrodes may however perturb the motion of the ions and displace them, which, in a magnetic gradient, directly transforms into magnetic field noise. 

We assume more generally an external force $\vect{F}(t) = e \vect{E}(t)$ where $e$ is the elementary charge and the electric field noise $\vect{E}(t) = (E_x(t), E_y(t), E_z(t))$ is characterized by its power spectral density $S_{E_i}(\omega) = \int^{+\infty}_{-\infty}R_{E_i}(\tau)e^{-i\omega\tau}d\tau$ with autocorrelation function $R(\tau) = \langle E_i(t) E_i(t+\tau) \rangle$. The trapped ion chain is modelled as a harmonic oscillator and the restoring force is therefore $\vect{F}(t) = - \vect{k}\vect{r}$, where $k_i = m\omega_i^2$ is the spring constant. We finally note the total motion of the ion $\vect{F} = -e \vect{\nabla} V(x, y, z, t)$, where $V$ is the potential created by the voltages on the electrodes \cite{wineland1998}. The total motion of the ion under the perturbative external force is retrieved by summing all forces. We limit ourselves to the axial direction of motion $z$ given that the confining potential is typically much weaker than in the radial direction, hence $\nu_z \ll \nu_{x,y}$, and the potential simplifies to $V(z) = z^2 \nu_z^2 m/2e$. The equation of motion becomes

\begin{equation} \label{eq:equation_of_motions}
m \ddot{z} + m \nu_z^2 z = e E_z(t),
\end{equation}

which is analogous to a forced harmonic oscillator. Taking the Fourier transform of equation (\ref{eq:equation_of_motions}) and using $S_z(\omega) = |\hat{z}(\omega)|^2$, the power spectral density of noise in the axial position is

\begin{equation} \label{eq:position_psd}
S_z(\omega) = \left[\frac{e}{m(\nu^2_z - \omega^2)}\right]^2 S_{E_z}(\omega).
\end{equation}

In a static magnetic gradient $\partial_z B$, a change in position $\Delta z$ results in a magnetic field variation $\Delta B = \partial_z B \Delta z$. Using this and relating electric field noise to voltage noise with $E_z(t) = \alpha_z V(t)/d$ where $\alpha_z$ is a geometric factor and $d$ is the ion to electrode distance, the power spectral density of magnetic field noise is found from equation (\ref{eq:position_psd}),

\begin{align} \label{eq:magnetic_noise_psd}
& S_B(\omega) = \left[\frac{\partial B}{\partial V}(\omega)\right]^2 S_V(\omega), \\
& \frac{\partial B}{\partial V}(\omega) = \frac{e \partial_zB \alpha_z}{m d (\nu^2_z - \omega^2)}. \label{eq:voltage_to_magnetic}
\end{align}

From equation (\ref{eq:voltage_to_magnetic}), one retrieves the scalings of the magnetic field noise with respect to the secular frequency ($\propto \nu_z^{-4}$) and the gradient ($\propto \partial_zB^2$). The geometric factor $\alpha_z$ reflects both the correlation of noise across the electrodes and their geometry. For perfectly correlated noise and a symmetric trap configuration, $\alpha_z = 0$. Note that for relevant parameters, $\nu_z^2 \gg \omega^2$ and the denominator of equation (\ref{eq:voltage_to_magnetic}) can therefore be replaced with $(\nu_z^2 - \omega^2) \rightarrow \nu_z^2$.

The continuous microwave drivings efficiently suppress qubit frequency noise arising from magnetic field noise. In a similar way to continuous dynamical decoupling in a 2-level system, the pair of dressing fields decouples the qubit from low frequency noise by opening an energy gap with frequency separation $\Omega_{dr}/\sqrt{2}$ and the system is only sensitive to noise near $S_B(\Omega_{dr}/\sqrt{2})$ \cite{cohen2015}. The coherence of the spin states follows an exponential decay $e^{-t \Gamma}$ where the decay rate is closely related to the magnetic field noise $\Gamma\sim S_B(\Omega_{dr}/\sqrt{2})$. Using the coherence decay as an approximate measure of fidelity, one therefore finds 

\begin{equation} \label{eq:fid_b_noise}
1 - \mathcal{F} \sim \tau S_B(\Omega_{dr}/\sqrt{2}),
\end{equation}

for a gate time $\tau = \pi/\sqrt{2}J_0$. Assuming that the magnetic field noise is dominated by voltage noise, the scalings of the fidelity (\ref{eq:fid_b_noise}) are governed by equation (\ref{eq:voltage_to_magnetic}) and the parameter ranges desired for the J-coupling gate become detrimental. For example, the J-coupling strength, and by extension the gate speed, increase quadratically with the magnetic field gradient ($\propto \partial_zB^2$). The magnetic field noise, however, also scales quadratically and therefore using a stronger gradient will not decrease errors from dephasing. Similarly, the gate speed is proportional to the secular frequency $\propto \nu_z^{-2}$. Recalling the scaling of the magnetic field noise $\propto \nu_z^{-4}$, decreasing the vibrational frequency will therefore increase the noise at a faster rate than the speed of the gate, which ultimately leads to higher infidelities.